\documentclass[a4paper,11pt]{article}
\usepackage[colorlinks=true,breaklinks=true,citecolor=blue,linkcolor=blue,urlcolor=blue,allcolors=blue]{hyperref}

\usepackage{graphicx}
\usepackage{amssymb,enumerate,graphicx,subfig,float,color,mathtools,amsmath,booktabs,multirow,array,xcolor,colortbl,stmaryrd,graphics}
\usepackage{authblk}
\usepackage[authoryear]{natbib}
\usepackage{bibentry}
\usepackage[section]{placeins} 
\usepackage{hyperref}

\newcommand{\amend}[1]{{#1}}

\newcommand{\changefourth}[1]{{{#1}}}
\newcommand{\tpmrevision}[1]{{{#1}}}
\newcommand{\tpmsecondrevision}[1]{{{#1}}}
\newcommand{\tpmsecondrevisionarticle}[1]{{\color{black}{#1}}}
\usepackage[font=footnotesize]{caption}
\newcommand{\norm}[1]{\left\lVert#1\right\rVert}
\usepackage[margin=2.5cm]{geometry}

\begin{document}

\title{\amend{Numerical investigation into coarse-scale models of\\ diffusion in complex heterogeneous media}}

%


\author{Nathan G. March$^{\text{a},}$\footnote{Corresponding author: \href{mailto:nathan.march@csiro.au}{nathan.march@csiro.au}.}\hspace{0.17cm}, Elliot J. Carr$^{\text{a}}$ and Ian W. Turner$^{\text{a},\text{b}}$\\
        \small $^{\text{a}}$School of Mathematical Sciences, Queensland University of Technology (QUT), Brisbane, Australia.\\
        \small $^{\text{b}}$ARC Centre of Excellence for Mathematical and Statistical Frontiers (ACEMS), Queensland University of Technology (QUT), Brisbane, Australia.}




\date{}

\maketitle

\begin{abstract}
Computational modelling of diffusion in heterogeneous media is prohibitively expensive for problems with fine-scale heterogeneities. A common strategy for resolving this issue is to decompose the domain into a number of non-overlapping sub-domains and homogenize the spatially-dependent diffusivity within each sub-domain (homogenization cell). This process yields a coarse-scale model for approximating the solution behaviour of the original fine-scale model at a reduced computational cost. In this paper, we study coarse-scale diffusion models in block heterogeneous media and investigate, for the first time, the effect that various factors have on the accuracy of resulting coarse-scale solutions. We present new findings on the error associated with homogenization as well as \tpmrevision{confirm via numerical experimentation that periodic boundary conditions are the best choice for the homogenization cell and demonstrate that the smallest homogenization cell that is computationally feasible should be used in numerical simulations.}
\end{abstract}

\section{Introduction}
\label{sec:introduction_paper_3}
\begin{figure}[h]
\centering
{\includegraphics[width=\textwidth]{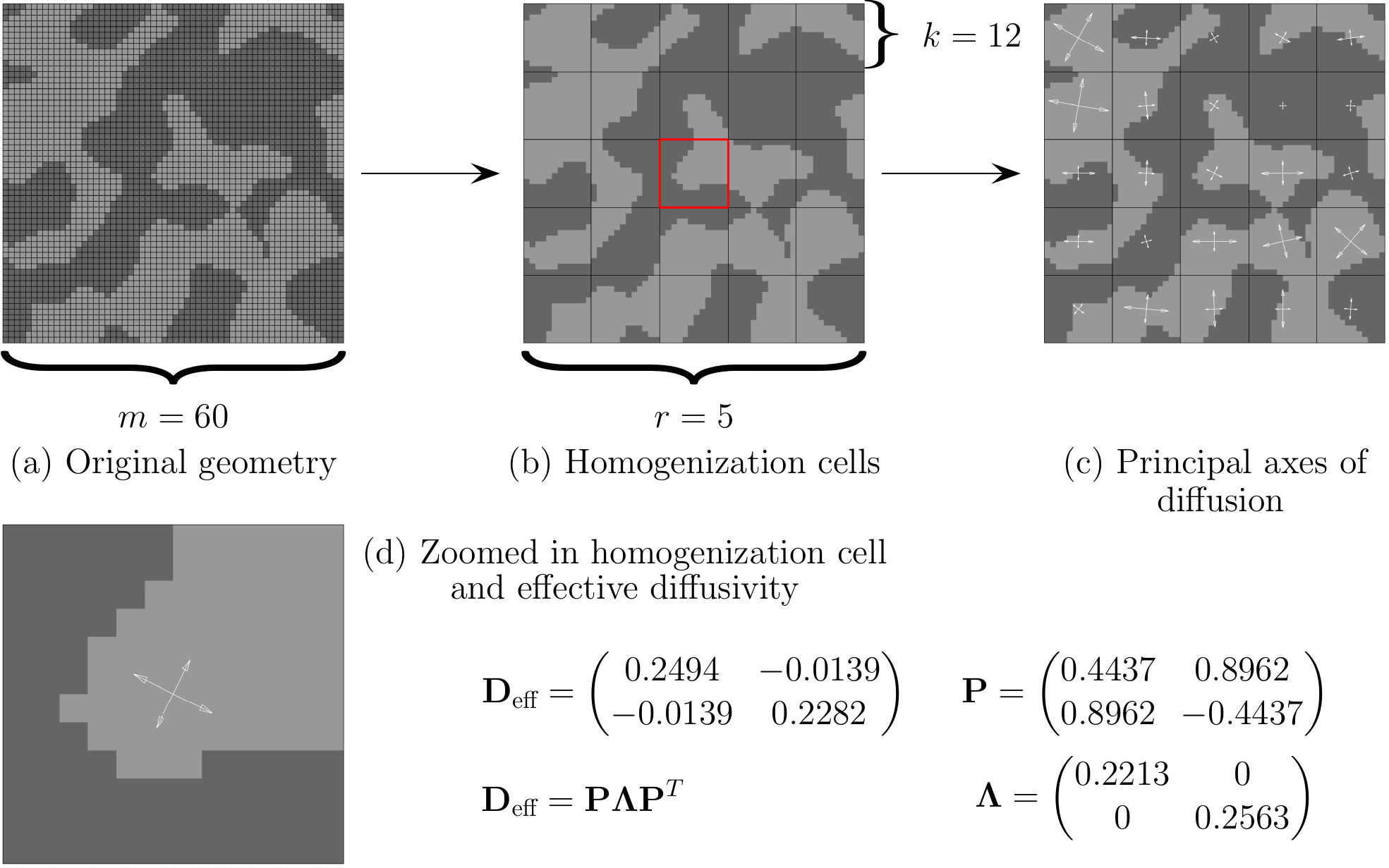}}
\caption{(a) Original geometry \tpmrevision{$\Omega$} consisting of an $m = 60$ by $m = 60$ array of blocks, where light grey blocks have a diffusivity of $1$ and dark grey blocks have a diffusivity of $0.1$. (b) The original $60$ by $60$ geometry has been divided into a $r = 5$ by $r = 5$ array of homogenization cells, where each homogenization cell is comprised of a $k = 12$ by $k = 12$ array of blocks. (c) In each of the $25$ homogenization cells, the principal axes of diffusion are overlaid as identified from the effective diffusivity for that cell. The lengths of the axes are scaled to reflect the diffusivity in the direction in which the axis points, where longer axes represent larger diffusivities. (d) A zoomed in view of the homogenization cell shown in red in (b), and the corresponding effective diffusivity. The effective diffusivity has been diagonalised to show the eigenvalues (diffusivities) and corresponding eigenvectors (principal axes of diffusion).}
\label{fig:figure1}
\end{figure}

Diffusive transport processes in heterogeneous media occur in a diverse range of environmental and industrial applications including fluid flow \citep{carr_2013,carr_2013b}, heat conduction \citep{hickson_2009a,hickson_2009b,carr_2016,carr_2018a,march_2019c}, contaminant transport in porous media \citep{liu_1998} and brain tumour growth \citep{asvestas_2014,mantzavinos_2014}. \tpmrevision{In this work, we consider the modelling of these processes via the steady-state diffusion equation:
\begin{align}
\label{eq:heterogeneous_transport}
0 =  \boldsymbol{\nabla}\cdot\left( D(\mathbf{x})\nabla u(\mathbf{x})\right), \quad \mathbf{x} \in \Omega,
\end{align}
in} which the diffusivity $D(\mathbf{x})$ varies spatially. In this work we consider a two-dimensional square domain $\Omega = [x_0,x_m] \times [y_0,y_m]$ and $\mathbf{x} = (x,y)$. If the scale at which the diffusivity $D(x,y)$ changes is small compared to the size of the domain $\Omega$, direct numerical solution of equation (\ref{eq:heterogeneous_transport}) \amend{is} computationally infeasible due to the very fine mesh required to capture the heterogeneity \citep{abdulle_2003,allaire_2005,arbogast_1993a,arbogast_1993b,carr_2016,chen_2008,davit_2013,jenny_2009}. This problem can be overcome by decomposing the domain $\Omega$ into a number of non-overlapping sub-domains (homogenization cells) and replacing the fine-scale diffusivity on each homogenization cell with a coarse-scale effective diffusivity \amend{(see Figure \ref{fig:figure1}).} 
In this paper, we study coarse-scale diffusion models in block heterogeneous media and investigate the effect that various factors have on the accuracy of resulting coarse-scale solutions. Recommendations are given for choosing the number of homogenization cells and the boundary conditions imposed on the homogenization cells to improve the match between coarse-scale and fine-scale solutions. \tpmrevision{We note that the methodology that we discuss in this work has similarities with a multigrid approach, as we use meshes of different levels of refinement in our numerical simulation, however we believe that this methodology is better classified as a partially-homogenized approach. We note that the methodology does not involve solving the diffusion equation (\ref{eq:heterogeneous_transport}) on multiple grids and instead involves a grid imposed on each \tpmsecondrevision{of the} homogenization cells (see Figure \ref{fig:figure1}(b)) and another grid imposed across the domain $\Omega$.}

Homogenization techniques have been considered by several authors including \citet{szymkiewicz_2012}, \citet{terada_2000} and \citet{vandersluis_2000}. \citet{szymkiewicz_2012} summarised three types of boundary conditions that can be imposed on the homogenization cell - periodic, confined and uniform (linear) (see Figure \ref{fig:BCs}). We consider these three types of boundary conditions in this work. \citet{terada_2000} compared the three types of boundary conditions in the modelling of stress and strain in heterogeneous materials and concluded that periodic boundary conditions were the best-performing conditions for both periodic and non-periodic media. They implemented a finite element method for simulation of the microstructure of their materials and compared the microscopic values for stress and strain to the homogenized values. \citet{kouznetsova_2001} compared microscale and macroscale simulations of the pure bending of porous aluminium and concluded that considering the microscale solution was computationally expensive due \amend{to} the large number of variables required to describe the microscale behaviour. They concluded that the homogenized model can estimate the macroscale behaviour, but only for small deformations to the medium. \citet{vandersluis_2000} considered the microstructural modelling of materials in order to determine their macroscopic mechanical properties. They considered periodic boundary conditions as well as mixed boundary conditions, in which one section of the boundary of the unit cell has the displacement prescribed and the other section of the boundary has the traction prescribed, and performed simulations using both uniform and irregular distributions of the microstructure materials. They concluded that the periodic boundary conditions were more appropriate than the mixed boundary conditions. \citet{sviercoski_2010} presented a method for approximating the effective diffusivity that uses a weighted average of the Cardwell and Parsons bounds \citep{cardwell_1945}. They used their calculations of the effective diffusivity to study diffusion through composite materials in which they conducted numerical simulations for three different steady-state diffusion problems. Each of these problems consisted of a two-dimensional rectangular domain with Dirichlet conditions on the left and right boundaries of the domain and homogeneous Neumann conditions on the top and bottom boundaries. They used a finite element method with triangular elements and computed a benchmark fine-scale solution and several coarse-scale solutions using homogenization cells of different sizes. These simulations investigated three different domains, all of different sizes and geometries and each simulation considered three or four different sizes of homogenization cell. They found that as the size of the homogenization cell quadruples (by doubling both the length and width of the cell), that the error between the fine-scale and coarse-scale solutions approximately doubles, which confirms theoretical results given by \citet{bensoussan_1978} and \citet{jikov_1994}. However, even though their estimate of the effective diffusivity as calculated using their weighted averaging method is accurate (or exact) for \tpmsecondrevision{most} homogenization cells, in some cases the effective diffusivities had errors in excess of $5\%$ as compared to an effective diffusivity calculated using the solution of a homogenization problem. Additionally, with only three simulations, it cannot be determined if similar results relating the size of the homogenization cell and the error would apply for other geometries.

After reviewing the available literature, we have identified \changefourth{the following} knowledge gaps that we aim to address in this work:
\begin{itemize}
\item \tpmrevision{Do periodic boundary conditions perform significantly better than confined and uniform conditions in the solution of heterogenenous diffusion problems?}
\item \amend{What information about the fine-scale solution is lost when the coarse-scale model is used instead?}
\item How does the homogenization error compare to the spatial discretisation error when solving the coarse-scale model numerically?
\item \tpmrevision{How much does the estimate of the average flux change when using the solution from the coarse-scale model compared to the fine-scale model?}
\item \tpmrevision{What effect (if any) do minor changes to the underlying geometry have on the solution obtained using the coarse-scale model?}
\end{itemize}

In order to answer these questions, we consider both fine-scale and coarse-scale diffusion models, which we discuss in section \ref{sec:diffusion_models}. The numerical methods we employ to solve the fine-scale and coarse-scale models and calculate the effective diffusivities are discussed in section \ref{sec:fvm}. In sections \ref{sec:results} and \ref{sec:qualitative_results} we implement some numerical experiments and discuss the results. Section \ref{sec:conclusion} concludes the paper.

\section{Diffusion problem}
\label{sec:diffusion_models}
In this section we define fully the diffusion problem considered in this paper. As stated earlier, we consider the square domain $\Omega = [x_0,x_m] \times [y_0,y_m]$ divided into an $m$ by $m$ grid of square blocks, where $m$ can be any positive integer. Each of these blocks may have a different diffusivity, however the diffusivity is constant across each block. Figure \ref{fig:figure1}(a) shows an example block domain $\Omega$ consisting of two materials with different diffusivities.

We consider both fine-scale and coarse-scale diffusion models. The key difference between these two models is that the fine-scale model uses an isotropic diffusivity $D(x,y)$ that varies rapidly across the domain and is described by equation (\ref{eq:heterogeneous_transport}), whereas the coarse-scale model uses a (possibly anisotropic) effective diffusivity  $\mathbf{D}_{\text{eff}}(x,y)$ that varies on a much coarser scale compared to the fine-scale diffusivity. The coarse-scale model is described by:
\begin{align}
\label{eq:PDE}
\tpmrevision{0}= \nabla \cdot (\mathbf{D}_{\text{eff}}(x,y) \nabla U(x,y)), \quad (x,y) \in \Omega,
\end{align}
where $U(x,y)$ is a coarse-scale approximation of $u(x,y)$ and $\mathbf{D}_{\text{eff}}(x,y) \in \mathbb{R}^{2 \times 2}$ is the effective diffusivity. In both the fine-scale and coarse-scale models we consider a general initial condition and on each of the four boundaries of the domain $\Omega$ we impose either a Dirichlet or homogeneous Neumann boundary condition. We detail our numerical solution strategy for both the fine-scale and coarse-scale models in section \ref{sec:fvm}.
\subsection{Calculation of effective diffusivity}
\begin{figure}[h]
\centering
\includegraphics[width=\textwidth,trim=1.1cm 1.1cm 1.1cm 0.5cm,clip]{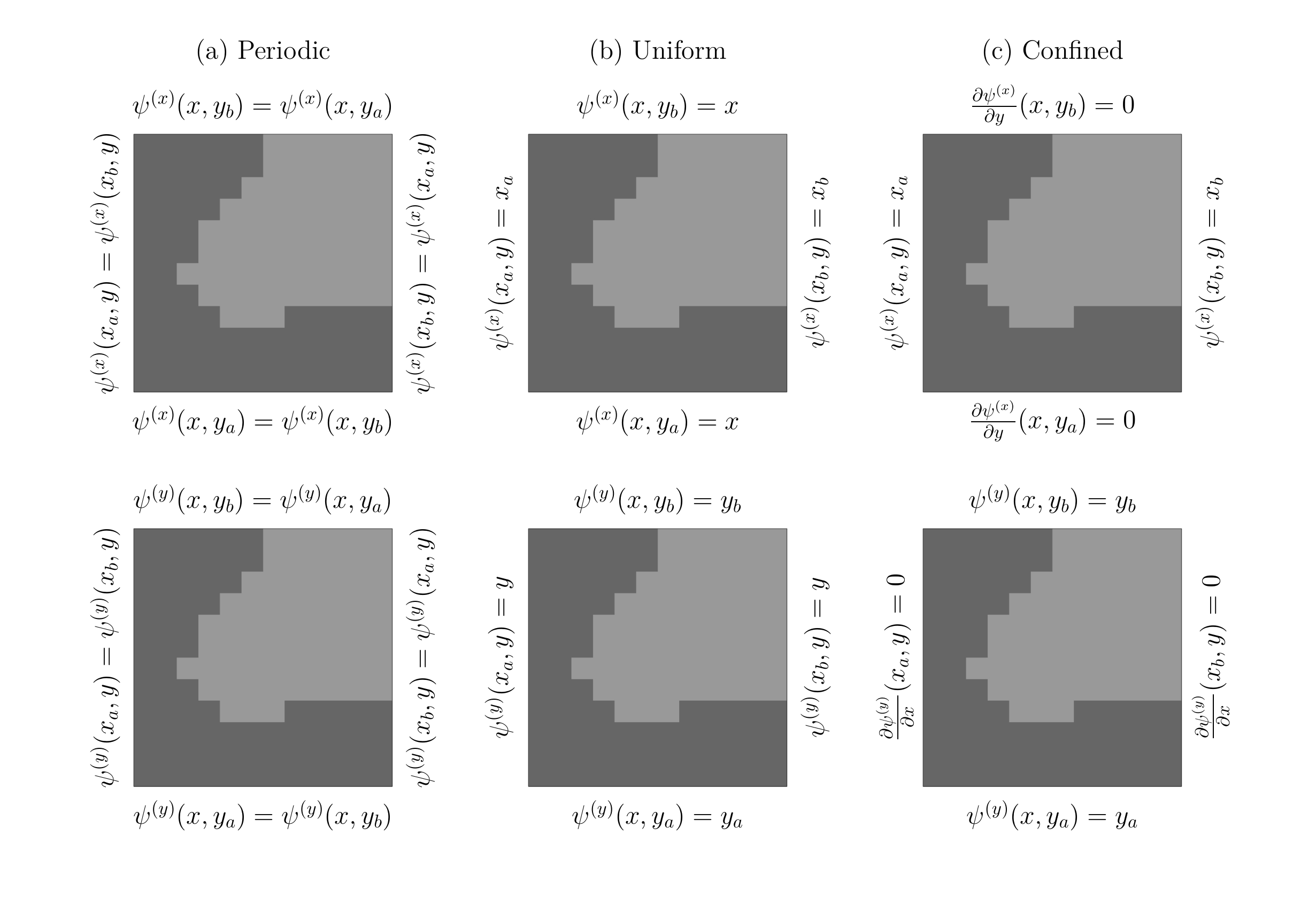}
\caption{(a)--(c) Boundary conditions for the periodic, uniform and confined problems with $\xi = x$ and $\xi = y$.} 
\label{fig:BCs}
\end{figure}

According to the homogenization literature \citep{hornung_1997,szymkiewicz_2005,davit_2013,carr_2013b,carr_2014,carr_2016b,carr_2017a}, the effective diffusivity $\mathbf{D}_{\text{eff}}(x,y)$ used in the coarse-scale model (\ref{eq:PDE}) can be calculated using the solution of an appropriate boundary value problem (or homogenization problem). In this work, we consider three common formulations for the homogenization problem involving periodic, confined and uniform boundary conditions (see Figure \ref{fig:BCs}) \citep{szymkiewicz_2012, renard_1997}. In this section, we discuss these three types of boundary conditions, their associated boundary value problems and how they are used to calculate the effective diffusivities for an arbitrary homogenization cell $\Omega_{C} = [x_{a},x_{b}]\times[y_{a},y_{b}]$ and in section \ref{sec:effective_fvm} we discuss our numerical solution strategy for solving these boundary value problems.

\subsubsection{Periodic boundary value problem}
The boundary value problem for periodic conditions consists of the following partial differential equation \citep{szymkiewicz_2012}:
\begin{gather}
\label{eq:diffusion_periodic}
0 = \nabla \cdot (D(x,y) \nabla(\psi^{(\xi)}(x,y)+\xi)), \quad (x,y) \in \Omega_C,
\end{gather}
where $\xi = x$ or $\xi = y$. The conditions ensuring the periodicity of the solution $\psi^{(\xi)}$ are shown in Figure \ref{fig:BCs}(a) and the following conditions ensure periodicity of the flux:
\begin{gather}
D(x_a,y)\frac{\partial}{\partial x}\left(\psi^{(\xi)}+\xi\right)\vline_{x = x_a}=D(x_b,y)\frac{\partial}{\partial x}\left(\psi^{(\xi)}+\xi\right)\vline_{x = x_b},\\
D(x,y_a)\frac{\partial}{\partial y}\left(\psi^{(\xi)}+\xi\right)\vline_{y = y_a}=D(x,y_b)\frac{\partial}{\partial y}\left(\psi^{(\xi)}+\xi\right)\vline_{y = y_b}.
\end{gather} 
Additionally, to \tpmsecondrevision{enforce} that the solution is unique, a zero mean condition is imposed:
\begin{align}
\label{eq:zero_mean}
\frac{1}{|\Omega_C|} \int_{y_{a}}^{y_b}\int_{x_{a}}^{x_b} \psi^{(\xi)}(x,y)\,dx\,dy = 0.
\end{align}
The first and second columns of the effective diffusivity are then calculated using the following integrals, respectively:
\begin{gather}
\label{eq:deff_formula_periodic_1}
[\mathbf{D}_{\mathrm{eff}}]_{(:,1)} = \frac{1}{|\Omega_C|}\int_{y_a}^{y_b}\int_{x_a}^{x_b} D(x,y) \nabla (\psi^{(x)}(x,y)+x)\, dx \, dy, \\
\label{eq:deff_formula_periodic_2}
[\mathbf{D}_{\mathrm{eff}}]_{(:,2)} = \frac{1}{|\Omega_C|}\int_{y_a}^{y_b}\int_{x_a}^{x_b} D(x,y) \nabla (\psi^{(y)}(x,y)+y)\, dx \, dy.
\end{gather}

\subsubsection{Uniform boundary value problem}
The boundary value problem for uniform conditions consists of the following partial differential equation \citep{szymkiewicz_2012}:
\begin{gather}
\label{eq:diffusion_uniform}
0 = \nabla \cdot (D(x,y) \nabla \psi^{(\xi)}(x,y)), \quad (x,y) \in \Omega_C,
\end{gather}
with the following boundary conditions imposed:
\begin{gather}
\hspace{-0.3cm}\psi^{(x)}(x_a,y) = x_a,\quad
\psi^{(x)}(x_b,y) = x_b,\quad
\psi^{(x)}(x,y_a) = x,\quad
\psi^{(x)}(x,y_b) = x,
\end{gather}
or
\begin{gather}
\hspace{-0.3cm}\psi^{(y)}(x_a,y) = y,\quad
\psi^{(y)}(x_b,y) = y,\quad
\psi^{(y)}(x,y_a) = y_a,\quad
\psi^{(y)}(x,y_b) = y_b.
\end{gather}
The first and second columns of the effective diffusivity are then calculated using the following integrals, respectively:
\begin{gather}
\label{eq:deff_formula_uniform_1}
[\mathbf{D}_{\mathrm{eff}}]_{(:,1)} = \frac{1}{|\Omega_C|}\int_{y_a}^{y_b}\int_{x_a}^{x_b} D(x,y) \nabla \psi^{(x)}(x,y)\, dx \, dy, \\
\label{eq:deff_formula_uniform_2}
[\mathbf{D}_{\mathrm{eff}}]_{(:,2)} = \frac{1}{|\Omega_C|}\int_{y_a}^{y_b}\int_{x_a}^{x_b} D(x,y) \nabla \psi^{(y)}(x,y)\, dx \, dy.
\end{gather}

\subsubsection{Confined boundary value problem}
The boundary value problem for confined conditions consists of the following partial differential equation \citep{szymkiewicz_2012}:
\begin{gather}
\label{eq:diffusion_confined}
0 = \nabla \cdot (D(x,y) \nabla \psi^{(\xi)}(x,y)), \quad (x,y) \in \Omega_C,
\end{gather}
which has the following boundary conditions imposed:
\begin{gather}
\psi^{(x)}(x_a,y) = x_a,\quad
\psi^{(x)}(x_b,y) = x_b,\quad
\frac{\partial \psi^{(x)}}{\partial y}\left(x,y_a\right) = 0,\quad
\frac{\partial \psi^{(x)}}{\partial y}\left(x,y_b\right)  = 0,
\end{gather}
or
\begin{gather}
\frac{\partial \psi^{(y)}}{\partial x}\left(x_a,y\right) = 0,\quad
\frac{\partial \psi^{(y)}}{\partial x}\left(x_b,y\right)  = 0,\quad
\psi^{(y)}(x,y_a) = y_a,\quad
\psi^{(y)}(x,y_b) = y_b.
\end{gather}
The first and second columns of the effective diffusivity are then calculated using identical integrals to those used for the \tpmrevision{uniform} conditions, equations (\ref{eq:deff_formula_uniform_1})--(\ref{eq:deff_formula_uniform_2}).
\section{Finite volume scheme}
\label{sec:fvm}
In this section we discuss the finite volume \amend{schemes} used to solve the fine-scale equation (\ref{eq:heterogeneous_transport}), coarse-scale equation (\ref{eq:PDE}) and equations used to calculate the effective diffusivity (\ref{eq:diffusion_periodic}), (\ref{eq:diffusion_uniform}) and (\ref{eq:diffusion_confined}). While we note that finite element methods are commonly used for such problems, particularly the calculation of effective diffusivities \citep{yi_2015,matache_2000,allaire_2005,talebi_2019,moulinec_1995,moulinec_1998,terada_2000,vandersluis_2000,kouznetsova_2001}, we use a finite volume method as we wish to build \tpmsecondrevision{on} and exploit the existing codes developed by the computational mathematics community at our university \tpmrevision{in recent decades} \citep{ferguson_1995,jayantha_2003}. \amend{The finite volume schemes used are based upon standard control volume finite element methods that mesh the domain $\Omega$ using square elements, such that the diffusivity is constant across each element, and approximate the solutions \tpmrevision{$u(x,y)$, $U(x,y)$} and $\psi^{(\xi)}(x,y)$ with a bilinear shape function, as has been used previously by \citet{moroney_2006, ferguson_1996} and \citet{sadrnejad_2012}.}

\subsection{Fine-scale model}
\label{sec:fine_scale}
\begin{figure}[t]
\centering
{\includegraphics[width=0.4\textwidth]{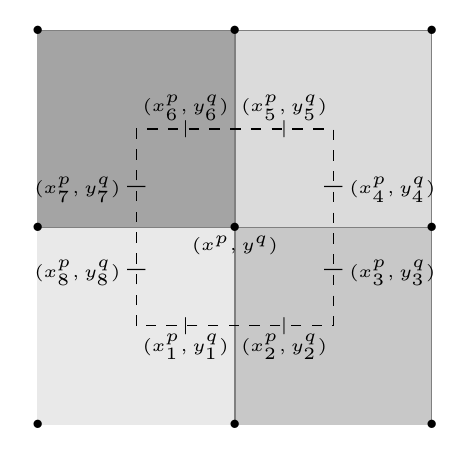}}
\caption{\amend{A control volume formed around the centre node $(x^{p},y^{q})$. The nine nodes form the vertices of four different elements, each of which has a constant diffusivity across the entire element. However, each element may have a different diffusivity, which is denoted by the different shading.}}
\label{fig:control_volume}
\end{figure}

\amend{We define a square mesh over the domain \amend{$\Omega = [x_0,x_m] \times [y_0,y_m]$} that consists of $N_f+1$ uniformly spaced nodes in both the $x$ and $y$ directions. The $x$-coordinates of the nodes are $x^p = x_0 + ph_f$ for $p = 0,\hdots,N_f$ and the $y$-coordinates are $y^q = y_0 + qh_f$ for $q = 0,\hdots,N_f$, where $h_f = (x_m-x_0)/N_f = (y_m-y_0)/N_f$. The spacing $h_f$ is chosen to ensure that nodes coincide with interfaces between blocks. We define the solution approximation at the node located at $(x^p,y^q)$ as \tpmrevision{$u^{p,q}$}. We decompose the domain into square elements by connecting the four nodes located at $(x^{p-1},y^{q-1}), (x^{p},y^{q-1}), (x^{p-1},y^{q})$ and $(x^{p},y^{q})$ for $p = 1,\hdots,N_f$ and $q = 1,\hdots,N_f$.}  We define a set of edges $\mathcal{E}_f^{p,q}$  (see Figure \ref{fig:control_volume}) around the node located at $(x^p,y^q)$ and connect them to form a control volume. The faces are constructed by connecting the centroid of an element to the midpoint of its edges \tpmrevision{(see Figure \ref{fig:control_volume})}. 

We integrate the governing partial differential equation (\ref{eq:heterogeneous_transport}) over the \amend{control volume} and apply the divergence theorem and midpoint rule to yield a finite volume equation. \amend{For a node located at $(x^p,y^q)$, except for those located on a boundary in which Dirichlet conditions are imposed, the finite volume equation takes the form:}
\begin{align}
\label{eq:FV3}
\tpmrevision{0}=    \sum_{s \in \mathcal{E}_f^{p,q}} \Phi_s^{p,q}, 
\end{align}
where $\Phi_s^{p,q}$ is the flux through the edge $s$ and is calculated as:
\begin{align}
\label{eq:flux1}
 \Phi_s^{p,q} = \begin{cases}  {D}(x_{s}^{p},y_{s}^{q}) \nabla \tpmrevision{u(x_{s}^{p},y_{s}^{q})} \cdot \mathbf{n}^{p,q}_{s}, \quad &\text{if} \quad (x_{s}^{p},y_{s}^{q}) \in \Omega,\\
 0, \quad &\text{if} \quad (x_{s}^{p},y_{s}^{q}) \in \partial\Omega,
\end{cases}
\end{align}
where $(x_s^p,y_s^q)$ is the midpoint of edge $s$ and $\mathbf{n}^{p,q}_{s}$ is the outward facing normal vector to edge $s$ with length $h_f/2$. We note that the flux corresponding to the edges located on the boundary is zero as the Neumann boundary conditions that we consider in this paper are homogeneous. For any node located on a boundary with Dirichlet conditions, the solution approximations corresponding to the nodes located along the boundary can be calculated directly, so no finite volume equation is required for these nodes. Any node adjacent to one of these boundary nodes is modified to account for these known solution values.

On each element, we approximate the solution with a bilinear interpolant $ \widetilde{u}(x,y)$. For an element \tpmrevision{$[x^{p-1},x^p] \times [y^{q-1},y^q]$}, the bilinear interpolant takes the form:
\begin{align}
\label{eq:bil_int}
\widetilde{u}(x,y) = a_1x + a_2y+a_3xy+a_4, \quad x^{p-1} < x < x^p, \quad y^{q-1} < y < y^q.
\end{align}
The coefficients $a_1,a_2,a_3$ and $a_4$ can be computed by noting that within each element, the solution values at the corners are known and by enforcing that the solution value as computed by the bilinear interpolant is equal to these values on each of the four corners, the following linear system can be generated:
\begin{align}
\begin{pmatrix}
x^{p-1} & y^{q-1} & x^{p-1}y^{q-1} & 1\\
x^{p} & y^{q-1} & x^{p}y^{q-1} & 1\\
x^{p-1} & y^{q} & x^{p-1}y^{q} & 1\\
x^{p} & y^{q} & x^{p}y^{q} & 1
\end{pmatrix}
\begin{pmatrix}
a_1 \\ a_2 \\ a_3 \\ a_4
\end{pmatrix} = 
\begin{pmatrix}
u^{p-1,q-1} \\ u^{p,q-1}\\ u^{p-1,q} \\ u^{p,q}
\end{pmatrix}.
\end{align}
This allows for the coefficients $a_1,a_2,a_3$ and $a_4$ to be expressed in terms of the solution approximations  $u^{p-1,q-1}, u^{p,q-1}, u^{p-1,q}$ and $u^{p,q}$ as:
\begin{align}
a_1 = \alpha_1 u^{p-1,q-1} + \alpha_2  u^{p,q-1} + \alpha_3  u^{p-1,q} + \alpha_4 u^{p,q},\\
a_2 = \alpha_5 u^{p-1,q-1} + \alpha_6  u^{p,q-1} + \alpha_7  u^{p-1,q} + \alpha_8 u^{p,q},\\
a_3 = \alpha_9 u^{p-1,q-1} + \alpha_{10}  u^{p,q-1} + \alpha_{11}  u^{p-1,q} + \alpha_{12} u^{p,q},\\
a_4 = \alpha_{13} u^{p-1,q-1} + \alpha_{14}  u^{p,q-1} + \alpha_{15}  u^{p-1,q} + \alpha_{16} u^{p,q},
\end{align}
where $\alpha_1,\hdots,\alpha_{16}$ can be computed in terms of the coordinates of the four corners of the element. This bilinear interpolant allows \amend{$\nabla \widetilde{u}(x,y) $ to be computed as:
\begin{align}
\label{eq:bil_flux}
\nabla \widetilde{u}(x,y)  = [a_1 + a_3y, a_2 + a_3x]^T.
\end{align}
This gradient is then used to approximate the flux term appearing in the finite volume equation (\ref{eq:FV3}).}

\subsection{Coarse-scale model}
\label{sec:coarse_scale}
The finite volume scheme for the coarse scale model is \amend{similar} to that of the fine-scale model  except a coarser mesh can be used as $\mathbf{D_{\text{eff}}}(x,y)$ varies at a coarser scale compared to $D(x,y)$.

We define a square mesh for the coarse-scale model over the domain \amend{$\Omega = [x_0,x_m] \times [y_0,y_m]$} that consists of $N_c+1$ uniformly spaced nodes in both the $x$ and $y$ directions. The $x$-coordinates of the nodes are $x^p = x_0 + ph_c$ for $p = 0,\hdots,N_c$ and the $y$-coordinates are $y^q = y_0 + qh_c$ for $q = 0,\hdots,N_c$, where $h_c = (x_m-x_0)/N_c = (y_m-y_0)/N_c$. The spacing $h_c$ is chosen to ensure that each control volume edge is located entirely within one homogenization cell. We define the solution at the node located at $(x^p,y^q)$ as \tpmrevision{$U^{p,q}$}. \tpmrevision{This solution is approximated on each element with a bilinear interpolant $\widetilde{U}(x,y)$. The remainder of the finite volume method for the coarse-scale model is implemented identically to the finite volume method for the fine-scale model.}

\subsection{\tpmrevision{Linear system for fine-scale and coarse-scale models}}
\tpmrevision{The finite volume equations for \amend{each} of the nodes \amend{for the fine-scale and coarse-scale models}  are collected to form a \tpmrevision{system of linear equations. For the fine-scale model, the linear system takes the form}:
\tpmsecondrevision{\begin{align}
\label{eq:linear_system}
\mathbf{A}_f\mathbf{u} = \mathbf{b}_f,
\end{align}
where}} $\mathbf{u}$ is a vector containing the discrete unknown values $u^{p,q}$ at each node, except for those corresponding to Dirichlet boundary conditions (as the value of the solution is known at these nodes), $\mathbf{A}_f$ is a matrix whose entries are determined using the finite volume equation (\ref{eq:FV3}) and the boundary conditions and $\mathbf{b}_f$ is a vector whose entries are determined by the boundary conditions. The solution of this linear system is \tpmsecondrevision{computed using the backslash operator in MATLAB}. Similarly, the linear system for the coarse-scale model takes the form:
\begin{align}
\label{eq:linear_system_coarse}
\mathbf{A}_c\mathbf{U} = \mathbf{b}_c,
\end{align}
where $\mathbf{U}$ is a vector containing the discrete unknown values $U^{p,q}$ at each node, except for those corresponding to Dirichlet boundary conditions and $\mathbf{A}_c$ and $\mathbf{b}_c$ are formed similarly to the fine-scale model. The solution of equation (\ref{eq:linear_system_coarse}) is \tpmsecondrevision{also} calculated \tpmsecondrevision{using the backslash operator in MATLAB.}

\subsection{Effective diffusivity}
\label{sec:effective_fvm}
We define a square mesh over the homogenization cell \amend{$\Omega_C = [x_a,x_b] \times [y_a,y_b]$} that consists of $N_e+1$ uniformly spaced nodes in both the $x$ and $y$ directions. The $x$-coordinates of the nodes are $x^p = x_a+ ph_e$ for $p = 0,\hdots,N_e$ and the $y$-coordinates are $y^q = y_0 + qh_e$ for $q = 0,\hdots,N_e$, where $h = (x_b-x_a)/N_e = (y_b-y_a)/N_e$. The spacing $h_e$ is chosen to ensure that nodes coincide with interfaces between blocks. The governing equation for the effective diffusivity, (\ref{eq:diffusion_periodic}),  (\ref{eq:diffusion_uniform}) or  (\ref{eq:diffusion_confined}), depending on \tpmsecondrevision{the chosen} boundary conditions, \amend{is} solved using the finite volume scheme for the fine-scale model with slight modifications. For periodic conditions, the finite volume equation for a node located at $(x^p,y^q)$ takes the form:
\begin{align}
\label{eq:FV_periodic}
0 =  \sum_{s \in \mathcal{E}_e^{p,q}}  {D}(x_{s}^p,y_{s}^q) (\nabla \psi^{(\xi)}(x_{s}^p,y_{s}^q)+\nabla \xi ) \cdot \mathbf{n}^{p,q}_{s}, 
\end{align}
where $(x_{s}^p,y_{s}^q)$ is the midpoint of edge $s$, $\mathcal{E}_e^{p,q}$ is the set of edges \tpmsecondrevision{surrounding} the node, $\mathbf{n}^{p,q}_{s}$ is the outward facing normal vector of edge $s$ with length $h_e/2$  and $\nabla \xi = [1,0]^T$ if $\xi = x$ and $\nabla \xi = [0,1]^T$ if $\xi = y$. The nodes along the boundaries are treated by modifying the finite volume equation (\ref{eq:FV_periodic}) to account for the periodic boundary conditions, as detailed by \citet{carr_2013b}. The zero mean condition (\ref{eq:zero_mean}) is accommodated by approximating the integrals with the midpoint rule, leading to the summation:
\begin{gather}
\label{eq:zero_mean_fvm}
\frac{h_e^2}{|\Omega_C|} \sum_{p = 0}^{N_e}\sum_{q = 0}^{N_e}\psi_{p,q}^{(\xi)}\ =0,
\end{gather}
where $\psi_{p,q}^{(\xi)} = \psi^{(\xi)}(x^p,y^q)$. For uniform and confined conditions, the finite volume equation for a node located at $(x^p,y^q)$, except for those located on a boundary in which Dirichlet conditions are imposed, takes the form:
\begin{align}
\label{eq:FV_uniform}
0 =   \sum_{s \in \mathcal{E}_e^{p,q}} \Phi_{s}^{p,q},
\end{align}
where:
\begin{align}
\label{eq:flux1}
\Phi_{s}^{p,q} = \begin{cases}  {D}(x_{s}^p,y_{s}^q) \nabla \psi^{(\xi)}(x_{s}^p,y_{s}^q) \cdot \mathbf{n}^{p,q}_{s}\quad &\text{if} \quad (x_{s}^p,y_{s}^q) \in \Omega_C,\\
 0, \quad &\text{if} \quad (x_{s}^p,y_{s}^q) \in \partial\Omega_C.
\end{cases}
\end{align}
We note that the flux corresponding to the edges located on the boundary is zero as the Neumann boundary conditions that occur in the confined conditions are homogeneous. For any node located on a boundary with Dirichlet conditions, the solution approximations corresponding to the nodes located along the boundary can be calculated directly, so no finite volume equation is required for these nodes. Any node adjacent to one of these boundary nodes is modified to account for these known solution values.

The finite volume equations for \amend{each} of the nodes \amend{for the effective diffusivity} are collected to form a \amend{system of linear equations} of the form:
\begin{align}
\label{eq:linear_system_ed}
\mathbf{A}_{e}\boldsymbol{\psi}^{(\xi)} = \mathbf{b}_{e},
\end{align}
where $\boldsymbol{\psi}^{(\xi)}$ is a vector containing the discrete unknown values $\psi^{(\xi)}_{p,q}$ at each node, except for those corresponding to Dirichlet boundary conditions, $\mathbf{A}_{e}$ is a matrix whose entries are determined using the finite volume equations (\ref{eq:FV_periodic})--(\ref{eq:FV_uniform}) and the boundary conditions and $\mathbf{b}_{e}$ is a vector whose entries are determined by the boundary conditions. \tpmsecondrevision{The solution of this linear system is calculated using the backslash operator in MATLAB.} The effective diffusivities are then calculated by approximating the integrals appearing in equations (\ref{eq:deff_formula_periodic_1})--(\ref{eq:deff_formula_periodic_2}) and (\ref{eq:deff_formula_uniform_1})--(\ref{eq:deff_formula_uniform_2}) with a midpoint rule, as was used previously by \citet{march_2019b}, leading to the following \tpmsecondrevision{approximations when periodic boundary conditions are imposed}:
\begin{gather}
\label{eq:deff_formula_periodic_1_sum}
[\mathbf{D}_{\mathrm{eff}}]_{(:,1)} = \frac{h_e^2}{|\Omega_C|}\sum_{p = 1}^{N_e}\sum_{q = 1}^{N_e}\ D(x_{\text{cen}}^p,y_{\text{cen}}^q) (\nabla \psi^{(x)}(x_{\text{cen}}^p,y_{\text{cen}}^q)+\mathbf{e}_1),\\
\label{eq:deff_formula_periodic_2_sum}
[\mathbf{D}_{\mathrm{eff}}]_{(:,2)} = \frac{h_e^2}{|\Omega_C|}\sum_{p = 1}^{N_e}\sum_{q = 1}^{N_e}\ D(x_{\text{cen}}^p,y_{\text{cen}}^q) (\nabla \psi^{(y)}(x_{\text{cen}}^p,y_{\text{cen}}^q)+\mathbf{e}_2),
\end{gather}
where $(x_{\text{cen}}^p,y_{\text{cen}}^q)$ is the centroid of the $(p,q)$th element, $\mathbf{e}_1 = [1,0]^T$ and $\mathbf{e}_2 = [0,1]^T$. The corresponding \tpmsecondrevision{approximations when uniform and confined boundary conditions are imposed} are:
\begin{gather}
\label{eq:deff_formula_uniform_1_sum}
[\mathbf{D}_{\mathrm{eff}}]_{(:,1)} = \frac{h_e^2}{|\Omega_C|}\sum_{p = 1}^{N_e}\sum_{q = 1}^{N_e}\ D(x_{\text{cen}}^p,y_{\text{cen}}^q) \nabla \psi^{(x)}(x_{\text{cen}}^p,y_{\text{cen}}^q),\\
\label{eq:deff_formula_uniform_2_sum}
[\mathbf{D}_{\mathrm{eff}}]_{(:,2)} = \frac{h_e^2}{|\Omega_C|}\sum_{p = 1}^{N_e}\sum_{q = 1}^{N_e}\ D(x_{\text{cen}}^p,y_{\text{cen}}^q) \nabla \psi^{(y)}(x_{\text{cen}}^p,y_{\text{cen}}^q).
\end{gather}

\section{Quantitative results}
\label{sec:results}
In this section we apply the fine-scale and coarse-scale models to a variety of different steady-state problems. The main purposes of this section are to understand to what extent the process of homogenization affects numerical simulations of diffusion across complex heterogeneous media; determine how to choose properties used \tpmsecondrevision{for} homogenization (such as the boundary conditions on the homogenization cell) that will generate a solution of the coarse-scale problem (\ref{eq:PDE}) that most closely resembles the solution of the corresponding fine-scale problem (\ref{eq:heterogeneous_transport}); to determine which aspects of the problem (such as the boundary conditions used in the fine-scale and coarse scale models) affect the error associated with homogenization\tpmsecondrevision{ and to understand what effect imposing a slight change to a geometry has on the solution of the coarse-scale model}.

To perform our numerical experiments, we consider 100 different heterogeneous geometries consisting of 60 by 60 blocks ($m=60$) generated using an algorithm presented in our previous work \citep{march_2019b}. This algorithm produces geometries with a fixed proportion of high diffusivity blocks that are aggregated together. For all tests, we consider the calculation of $\mathbf{D}_{\text{eff}}$ using the periodic, uniform and confined boundary conditions on the homogenization cell discussed in section \ref{sec:diffusion_models}. We consider varying levels of homogenization characterised by \amend{a homogenization parameter $k$, defined as follows}. If an $m$ by $m$ domain is homogenized with homogenization parameter $k$, then the domain is split into \amend{$r^2 = (m/k)^2$} homogenization cells, each of which is comprised of a $k$ by $k$ array of identically sized blocks (see Figure \ref{fig:figure1}). On each of these \amend{$r^2$} homogenization cells, the effective diffusivity is computed and that effective diffusivity is used for the entirety of the homogenization cell. For a particular domain, any $k$ that divides $m$ is a valid homogenization parameter. The homogenization parameter $k = 1$ corresponds to the fine-scale problem, in which no effective diffusivity is calculated and $k = m$ is the fully homogenized problem, in which an effective diffusivity valid across the entire domain is computed. It is expected that as $k$ increases, the error relative to the fine-scale solution should increase, as each effective diffusivity is valid across a larger portion of the domain. \tpmsecondrevision{The first error metric that we use to assess the accuracy of the coarse-scale solution for different homogenization parameters, mesh sizes and boundary conditions is a solution-based error defined as:}
\tpmsecondrevision{\begin{align}
\label{eq:error}
E_G^U = \frac{\norm{\mathbf{U} - \mathbf{u}}_2}{ \norm{{\mathbf{u}}}_2}.
\end{align}
In this analysis we consider the mean error over 100 different geometries:}
\begin{align}
\label{eq:mean_error}
\tpmrevision{E^U} = \frac{1}{100}\sum_{G = 1}^{100} \tpmrevision{E_G^U}.
\end{align}
For all tests, we consider the following two different combinations of either Dirichlet or Neumann boundary conditions:
\begin{align}
\label{eq:coarse_BC1}
&\text{BC1:} \quad U(x_0,y) = 1,\quad  U(x_m,y) = 0,\quad  \frac{\partial U(y_m,x)}{\partial y} = 0, \quad  \frac{\partial U(y_0,x)}{\partial y} = 0,\\
\label{eq:coarse_BC3}
& \text{BC2:}\quad \frac{\partial U(x_0,y)}{\partial x} = 0,\quad \frac{\partial U(x_m,y)}{\partial x} = 0,\quad U(y_m,x) = 0,\quad U(y_0,y) = 1.
\end{align}
We choose these boundary conditions so that the observed solution behaviour is due to the heterogeneity of the geometry and not the coarse-scale boundary conditions.

All tests are completed using geometries consisting of two different media, one with diffusivity $D_1$ and the other with diffusivity $D_0$. We define the volume fraction $\varepsilon_1 = A_1/(A_1+A_0)$ where $A_1$ and $A_0$ are the areas of media with diffusivities $D_1$ and $D_0$, respectively. We also define the diffusivity ratio as $\varepsilon_2 = D_1/D_0$. In both \tpmsecondrevision{sections \ref{sec:results} and \ref{sec:qualitative_results}} the benchmark fine-scale solutions are computed using a mesh with $N_f = m$ and the effective diffusivities corresponding to homogenization parameter $k$ are computed using a mesh with $N_e = 3k$.

\subsection{Test 1: Relative errors of problems computed using a constant mesh}
\label{sec:test1}
\begin{figure}[h]
\centering
\subfloat[BC1]{\includegraphics[width=0.45\textwidth]{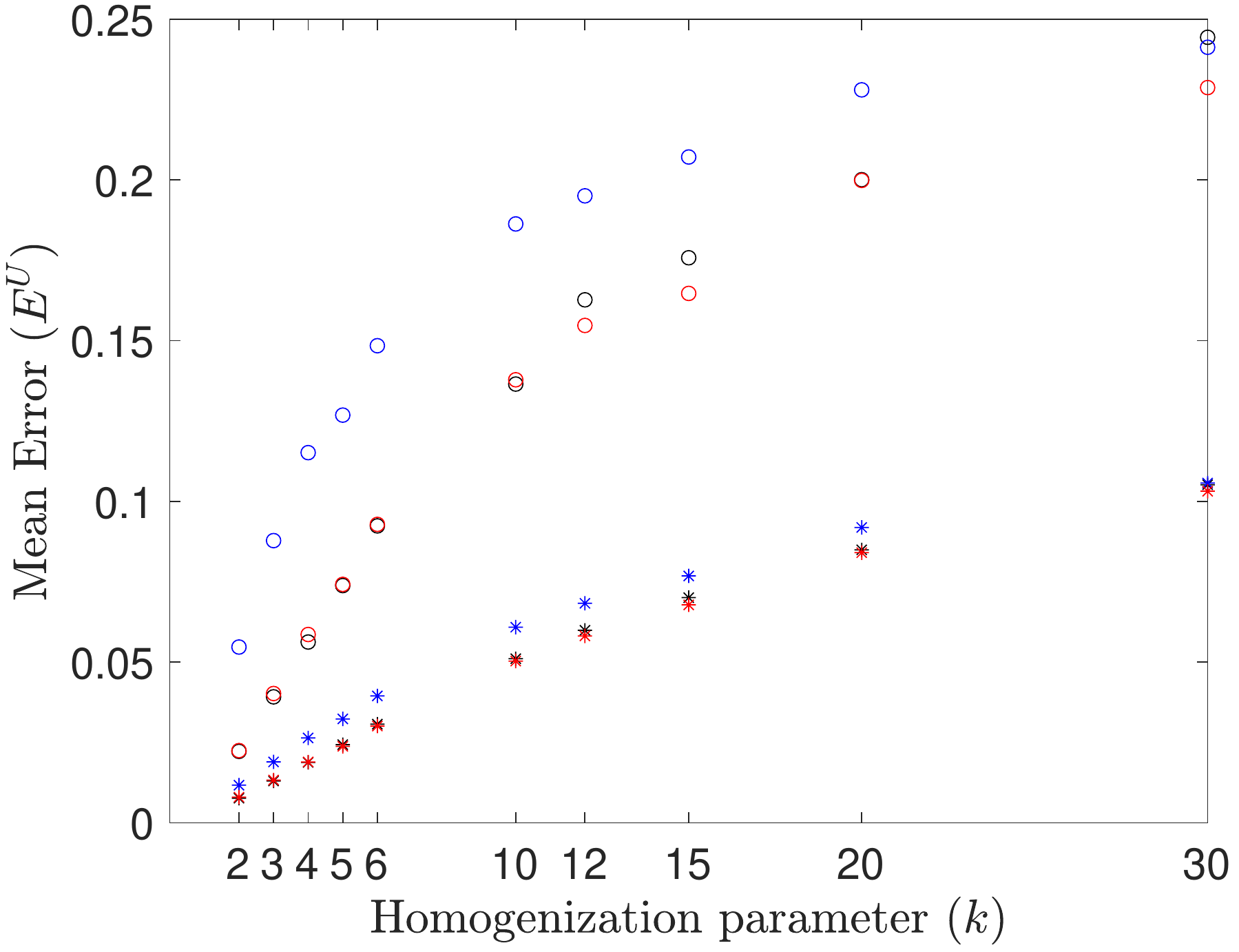}}\hspace{0.05\textwidth}
\subfloat[BC2]{\includegraphics[width=0.45\textwidth]{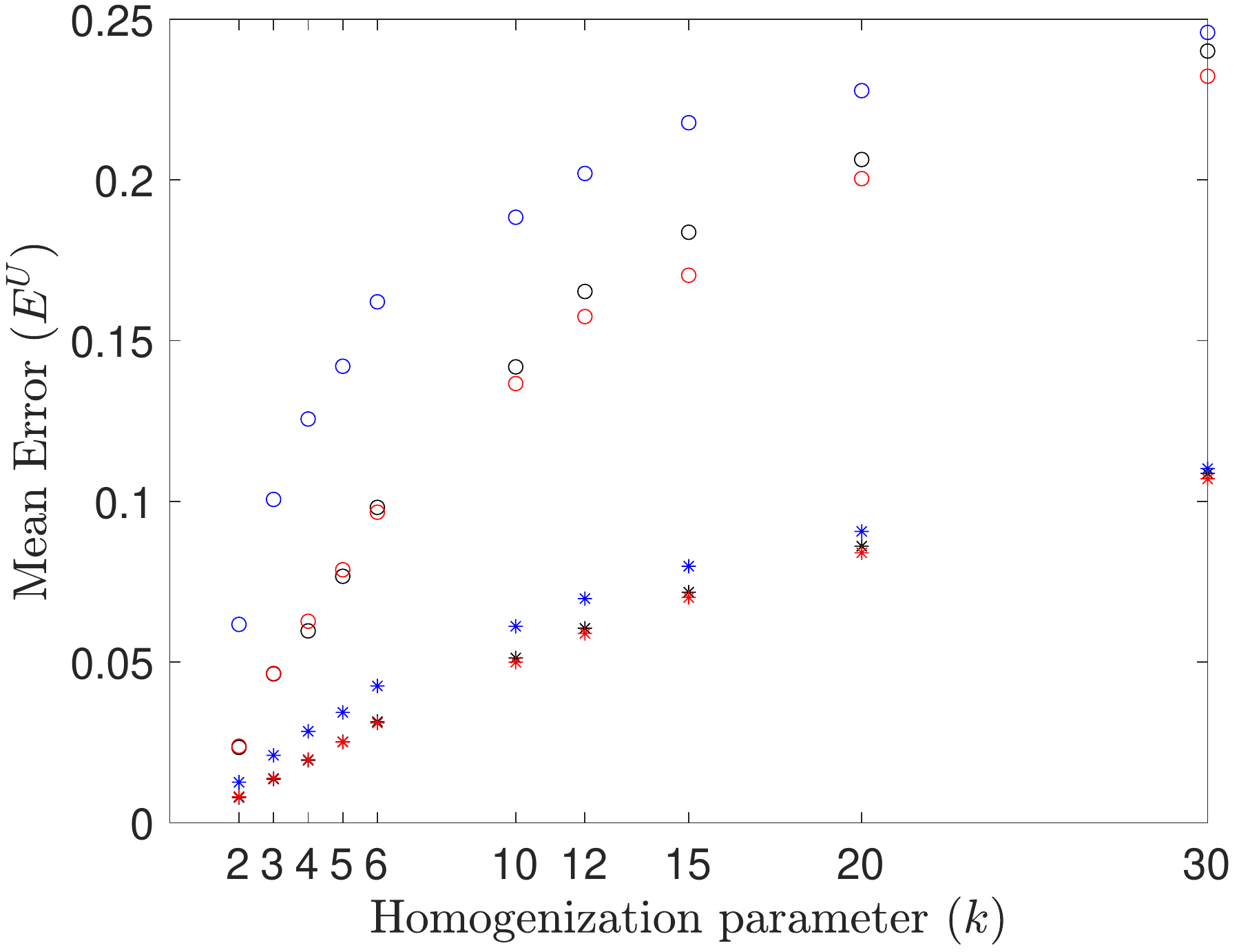}}
\caption{\tpmrevision{Mean \tpmsecondrevision{solution-based} error of the coarse-scale approximation $\mathbf{U}$ vs homogenization parameter for 100 geometries generated with a volume fraction of $\varepsilon_1 =50\%$ using periodic (black markers), uniform (\textcolor{blue}{blue} markers) and confined (\textcolor{red}{red} markers) boundary conditions. The stars $(*)$ represent diffusivity ratios $\varepsilon_2 =10$ and the circles $(\circ)$ represent $\varepsilon_2 =100$.}}
\label{fig:vf_50}
\end{figure}

{In this test we consider the previously mentioned 100 different geometries with volume fractions $\varepsilon_1$ of $25\%, 50\%$ and $75\%$ and diffusivity ratios $\varepsilon_2$ of $10$ and $100$.  We consider the homogenization parameters \tpmrevision{$k = 1,2,3,4,5,6,10,12,15,20,30$} and all three types of boundary conditions on the homogenization cell. The mesh size for the coarse-scale model is set to be \tpmrevision{$N_c = N_f$}. Preliminary results did not \tpmsecondrevision{show} significant differences in errors across different volume fractions $\varepsilon_1$, so only the results for $\varepsilon_1 = 50\%$ are shown in Figure \ref{fig:vf_50}. For the diffusivity ratio $\varepsilon_2 = 10$ and homogenization parameters $k\leq20$, the periodic and confined boundary conditions performed \tpmsecondrevision{similarly} and were more accurate than the uniform boundary conditions. For the \tpmrevision{homogenization parameters} $k>20$, all three boundary conditions \tpmsecondrevision{yielded similar results}. In the case of diffusivity ratio $\varepsilon_2 = 100$ and homogenization parameters $k\leq15$, the periodic and confined boundary conditions performed approximately equally and were significantly more accurate than the uniform boundary conditions. For $15 \leq k \leq 30$, the confined boundary conditions performed slightly better than the periodic boundary conditions, which performed  better than the uniform boundary conditions. \tpmrevision{As both plots in Figure \ref{fig:vf_50} look similar, the different coarse-scale boundary conditions (\ref{eq:coarse_BC1})--(\ref{eq:coarse_BC3}) appear to have little effect on the performance of the coarse-scale model.}}

\tpmrevision{We also compare the average flux calculated using the coarse-scale model to the average flux calculated using the fine-scale model. To calculate the average flux for the fine-scale model, we first calculate the gradient $\nabla\widetilde{u}(x,y)$ at the centroid of the element $[x^{p-1},x^p] \times [y^{q-1},y^q]$ using the finite volume scheme discussed in section \ref{sec:fvm} by evaluating equation (\ref{eq:bil_flux}) at the centroid of the element $(x_{\text{cen}}^p,y_{\text{cen}}^q)$ for $p = 1,\hdots,N_f$ and $q = 1,\hdots,N_f$. We then calculate the flux vector $\mathbf{J}_f^{p,q}$ as:
\begin{align}
\label{eq:flux_vector_fine}
\mathbf{J}_f^{p,q} = D(x_{\text{cen}}^p,y_{\text{cen}}^q) \nabla\widetilde{u}(x_{\text{cen}}^p,y_{\text{cen}}^q), 
\end{align}
for $p = 1,\hdots,N_f$ and $q = 1,\hdots,N_f$. The flux vector for the coarse-scale model, $\mathbf{J}_c^{p,q}$, is calculated similarly:
\begin{align}
\label{eq:flux_vector_coarse}
\mathbf{J}_c^{p,q} = D(x_{\text{cen}}^p,y_{\text{cen}}^q) \nabla\widetilde{U}(x_{\text{cen}}^p,y_{\text{cen}}^q), 
\end{align}
for $p = 1,\hdots,N_c$ and $q = 1,\hdots,N_c$. We note that we use the fine-scale diffusivity $D(x,y)$ in the calculation of the flux corresponding to the coarse-scale model, rather than the effective diffusivity $\mathbf{D}_{\text{eff}}(x,y)$. In preliminary testing, we used $\mathbf{D}_{\text{eff}}(x,y)$ in equation (\ref{eq:flux_vector_coarse}), rather than the fine-scale diffusivity $D(x,y)$, however we found that the estimates of the flux vector were more accurate (as compared to the flux calculated using equation (\ref{eq:flux_vector_fine})) when $D(x,y)$ is used. We note that using $D(x,y)$ in the calculation of the flux for the coarse-scale model is reasonable, as $D(x,y)$ is known everywhere in the domain $\Omega$ and is used to calculate the effective diffusivity $\mathbf{D}_{\text{eff}}(x,y)$. We define the matrices $\mathbf{J}_f \in \mathbb{R}^{2 \times N_f^2}$ and $\mathbf{J}_c \in \mathbb{R}^{2 \times N_c^2}$, which contain the components of the flux vectors $\mathbf{J}_f^{p,q}$ and $\mathbf{J}_c^{p,q}$, respectively, on each element. To calculate the average flux vectors, we compute the mean of the $N_f^2$ and $N_c^2$ columns of $\mathbf{J}_f$ and  $\mathbf{J}_c$, denoted as $\mathbf{\overline{J}}_f \in \mathbb{R}^{2 \times 1}$ and  $\mathbf{\overline{J}}_c \in \mathbb{R}^{2 \times 1}$, respectively. \tpmsecondrevision{The second error metric that we use is the flux-based error, defined as the error in the average flux between the coarse-scale and fine-scale solutions:}
\tpmsecondrevision{\begin{align}
\label{eq:flux_error}
\mathbf{E}_G^J = \, \vline{\mathbf{\overline{J}}_c - \mathbf{\overline{J}}_f\vline} \varoslash  \vline \mathbf{\overline{J}}_f \vline,
\end{align}
where $\varoslash$ represents Hadamard/element-wise division and $\vline  \,\cdot \, \vline$ indicates absolute value}. We then average this error over the $100$ different geometries:
\begin{align}
\label{eq:mean_error}
\mathbf{E}^J = \frac{1}{100}\sum_{G = 1}^{100} \mathbf{E}_G^J.
\end{align}
We note that $\mathbf{E}^J \in  \mathbb{R}^{2 \times 1}$ and that the first component of $\mathbf{E}^J$ represents the average error of the flux in the $x$-direction and that the second component represents the average error of the flux in the $y$-direction. However, we consider only the first component $\mathbf{E}^J_{[1]}$ in the analysis for the BC1 (\ref{eq:coarse_BC1}) boundary conditions and second component $\mathbf{E}^J_{[2]}$ in the analysis for the BC2 (\ref{eq:coarse_BC3}) boundary conditions, as the flux in the $y$-direction for BC1 boundary conditions and the flux in the $x$-direction for BC2 boundary conditions is very small, which can cause large relative errors in the components of $\mathbf{E}^J$ calculated using these fluxes.

\begin{figure}[h]
\centering
\subfloat[BC1]{\includegraphics[width=0.45\textwidth]{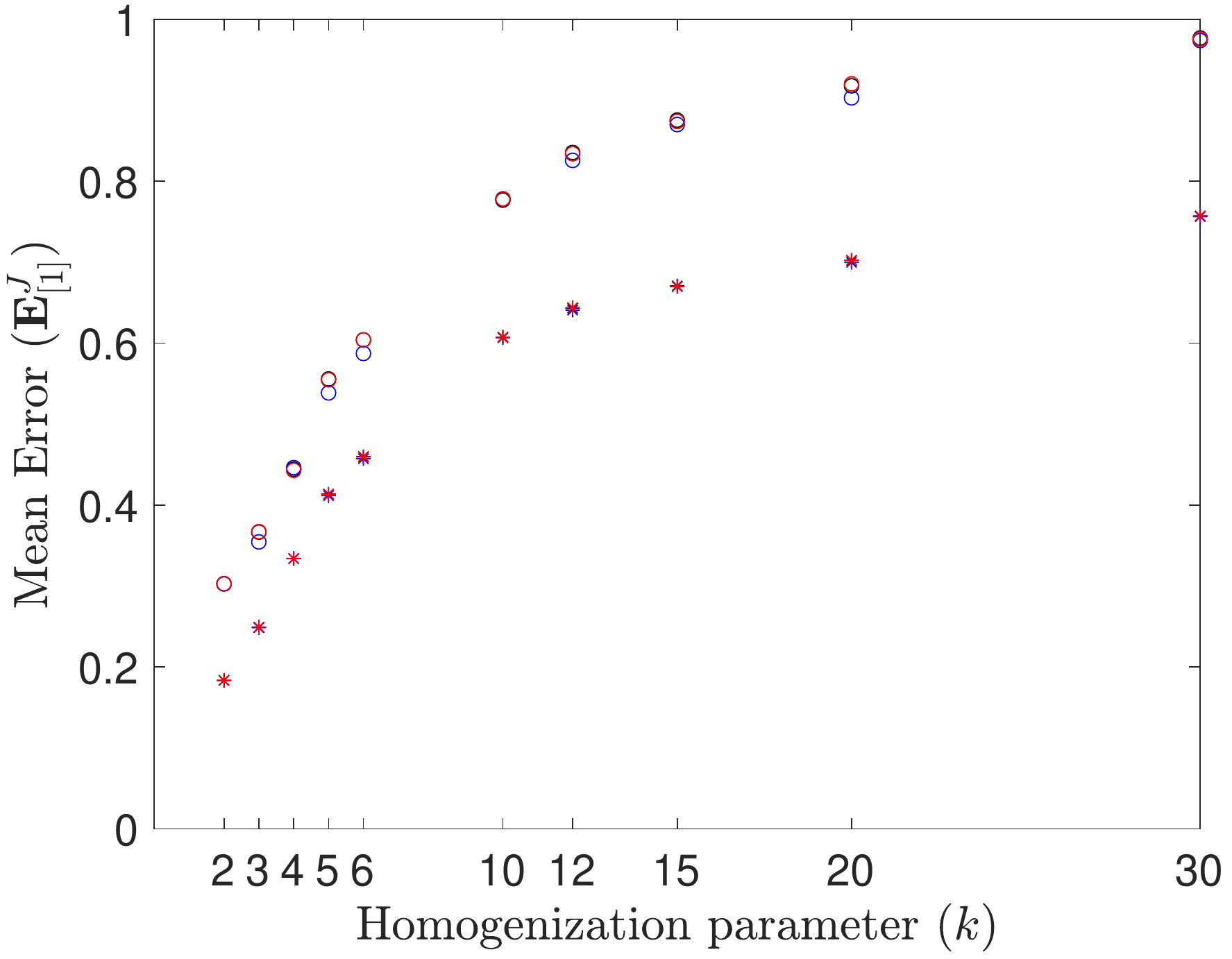}}\hspace{0.05\textwidth}
\subfloat[BC2]{\includegraphics[width=0.45\textwidth]{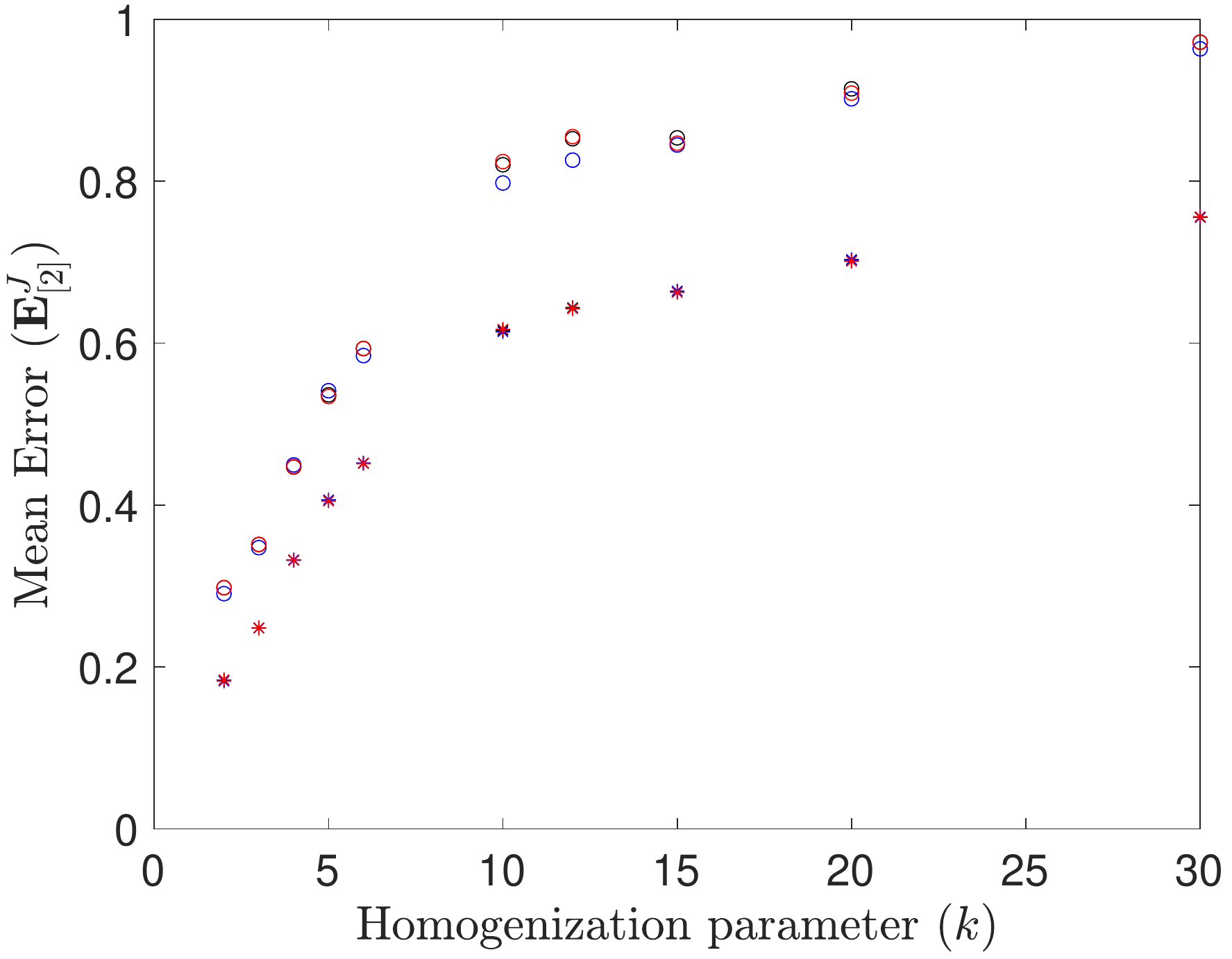}}
\caption{\tpmrevision{\tpmsecondrevision{Mean flux-based error of the coarse-scale approximation $\mathbf{U}$} vs homogenization parameter for 100 geometries generated with a volume fraction of $\varepsilon_1 =50\%$ using periodic (black markers), uniform (\textcolor{blue}{blue} markers) and confined (\textcolor{red}{red} markers) boundary conditions. The stars $(*)$ represent diffusivity ratios $\varepsilon_2 =10$ and the circles $(\circ)$ represent $\varepsilon_2 =100$.}}
\label{fig:vf_50_flux}
\end{figure}
As seen in Figure \ref{fig:vf_50_flux}, for both BC1 and BC2 conditions, all three boundary conditions \tpmsecondrevision{performed  comparably in terms of accuracy} for the diffusivity ratio $\varepsilon_2 = 10$ and that the periodic and confined conditions \tpmsecondrevision{performed similarly} for $\varepsilon_2 = 100$, with uniform conditions performing slightly better for some values of $k$. Similarly to Figure \ref{fig:vf_50}, the \tpmsecondrevision{flux-based error} increases as $k$ increases, from approximately $0.2$ to $0.75$ for $\varepsilon_2 = 10$ and $0.3$ to $1$ for $\varepsilon_2 = 100$. We also note that the \tpmsecondrevision{flux-based error shown} in Figure \ref{fig:vf_50_flux} is much higher than the \tpmsecondrevision{solution-based error shown} in Figure \ref{fig:vf_50}. We believe that the larger \tpmsecondrevision{flux-based error} is due to the the tendency of coarse-scale solutions to ``smoothen" over fine-scale details (see Figure \ref{fig:overlaid_axes_steady_state_periodic_BC_constant_mesh}), which can cause the flux calculated using the coarse-scale solution to differ significantly from the flux calculated using the fine-scale solution.}

\subsection{Test 2: Relative errors of problems computed using a varying mesh}
\label{sec:test2}
\amend{In this test, we consider the same 100 different geometries from Test 1. We consider only the homogenization parameters $k = 1,2,3,4,5,6$ and all three types of boundary conditions on the homogenization cell. The mesh size for the coarse-scale model is set to be $N_c = r $, which is the coarsest mesh that allows for each element to be contained entirely within homogeneous material. This differs from Test 1, as the simulations in Test 2 use a coarser mesh. As the mesh size varies with $k$, to calculate the \tpmsecondrevision{solution-based error} (\ref{eq:error}), bilinear interpolation is used to interpolate the coarse-scale solution onto the fine-scale mesh. The choices of $k$ are considered only up to $k = 6$, as opposed to the \tpmsecondrevision{$k = 30$ used} in Test 1 as the larger $k$ values correspond to very coarse meshes. To compare the results for Test 2 against those for Test 1, we calculate the error ratio $R = E_{\text{var}}/E_{\text{con}}$, where $E_{\text{var}}$ is the mean error (\ref{eq:mean_error}) as calculated using the varying mesh and $E_{\text{con}}$ is the mean error (\ref{eq:mean_error}) as calculated using the constant mesh.}
 
\begin{figure}[h]
\centering
\subfloat[BC1]{\includegraphics[width=0.45\textwidth]{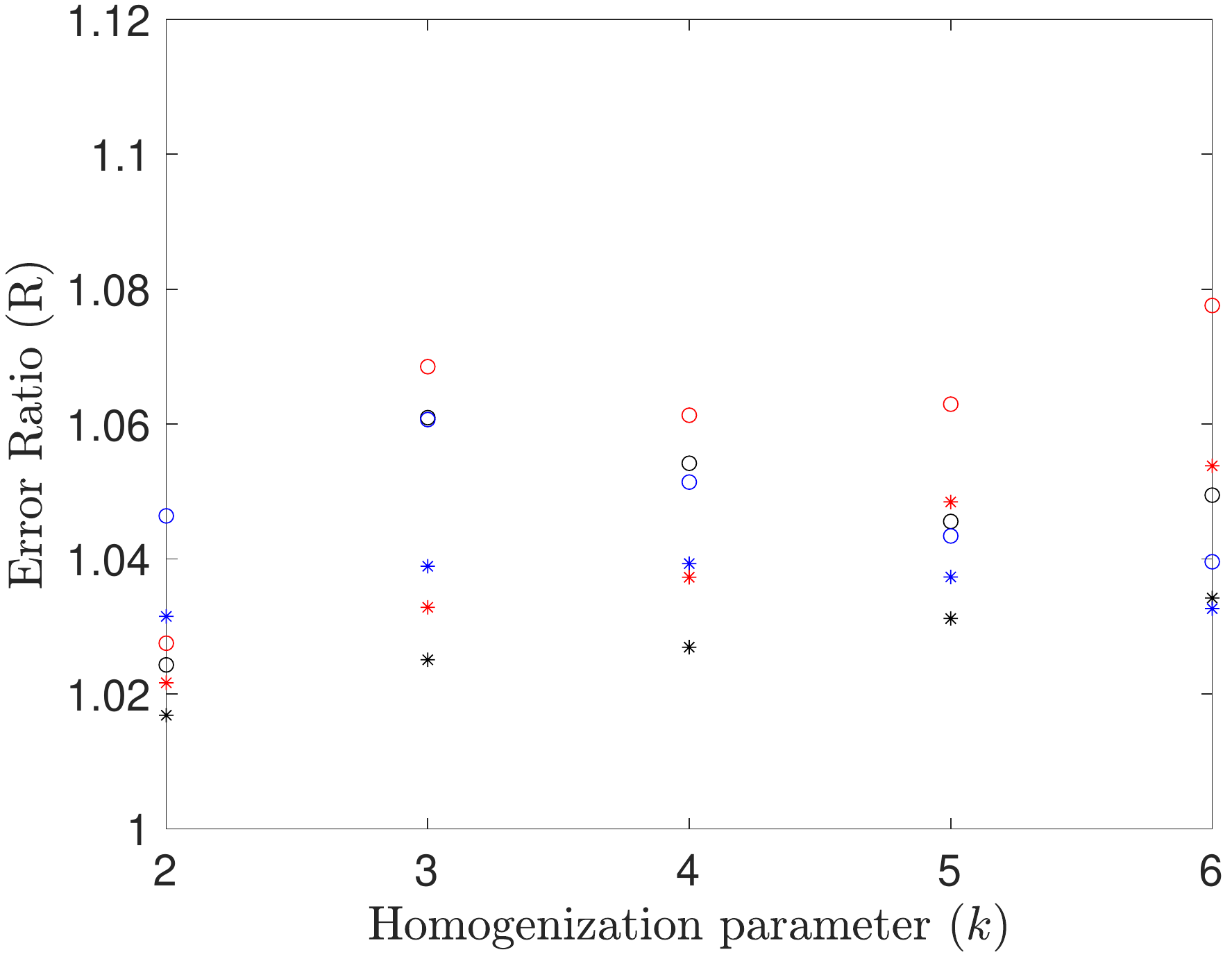}}\hspace{0.05\textwidth}
\subfloat[BC2]{\includegraphics[width=0.45\textwidth]{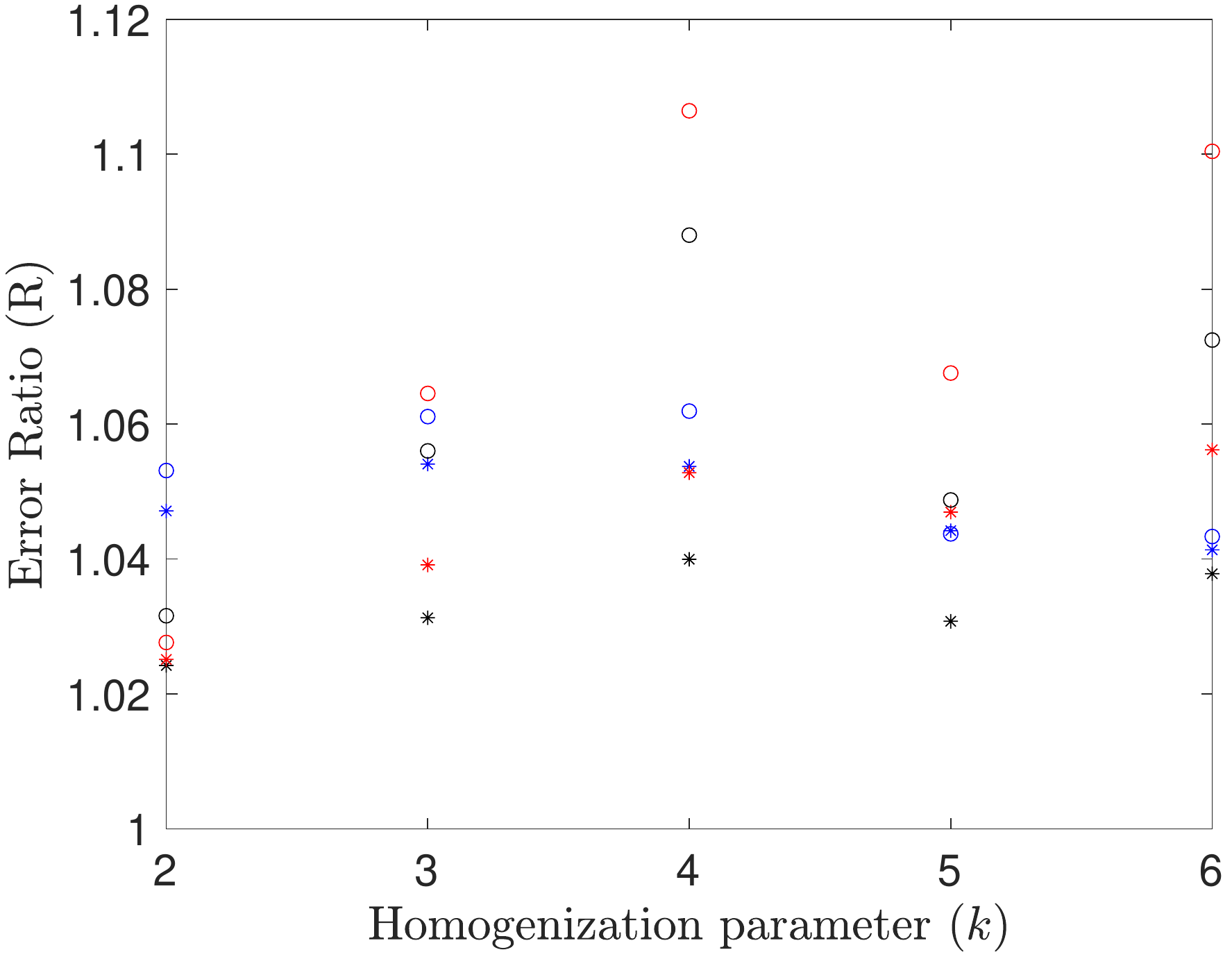}}
\caption{Error ratio ($R$) vs homogenization parameter ($k$) for 100 geometries generated with a volume fraction of $\varepsilon_1 =50\%$  using periodic (black markers), uniform (\textcolor{blue}{blue} markers) and confined (\textcolor{red}{red} markers) boundary conditions. The stars $(*)$ represent diffusivity ratios $\varepsilon_2 =10$ and the circles $(\circ)$ represent $\varepsilon_2 =100$.}
\label{fig:vf_50_dr_10_er}
\end{figure}

\amend{Similarly to Test 1, preliminary results did not show significant differences in errors across different volume fractions $\varepsilon_1$, so only the results for $\varepsilon_1 = 50\%$ are shown in Figure \ref{fig:vf_50_dr_10_er}. The results varied significantly for different combinations of coarse-scale boundary conditions. The purpose of this test is to determine what effect the use of a coarser mesh has on the relative error. We expect that the error would increase with a coarser mesh, as the spatial discretisation error associated with finite volume methods would generally increase as the number of nodes decreases. \tpmrevision{For both coarse-scale boundary conditions, the relative errors for diffusivity ratios $\varepsilon_2 = 10$ were generally lower than the corresponding relative errors for diffusivity ratios $\varepsilon_2 = 100$.}}

\tpmrevision{For BC1 (Figure \ref{fig:vf_50_dr_10_er}(a)), the error ratio is between approximately $1.02$ and $1.08$ for any diffusivity ratio or volume fraction and for BC2 (Figures \ref{fig:vf_50_dr_10_er}(b)), the error ratio is between approximately $1.02$ and $1.1$. There is no clear relationship between the homogenization parameter $k$ and the error ratio for any choices of coarse-scale boundary conditions, boundary conditions on the homogenization cell, volume fraction $\varepsilon_1$ or diffusivity ratio $\varepsilon_2$. However as the error ratio is between $1.02$ and $1.1$, the spatial discretisation error introduced through the use of a coarser mesh is between $2\%$ and $10\%$ of the homogenization error and thus the homogenization error is approximately $10$ to $50$ times larger than the spatial discretisation error associated with our implemented finite volume method.}

\subsection{Recommendations for homogenizing block heterogeneous media}
\label{sec:heuristics}
Based on the results generated and presented in section \ref{sec:results} and other properties of the different boundary conditions for the homogenization cell, we make some recommendations for homogenizing block heterogeneous media. We assume that for a given problem, the coarse-scale boundary conditions, initial condition, $m$ and $D_{i,j}$ for all $i = 1,\hdots,m$ and $j = 1,\hdots,m$ are all known. We also assume that there exists a value $N_{\text{max}}$, such that a mesh comprised of $N_{\text{max}}+1$ uniformly spaced nodes in both the $x$ and $y$ directions is the most refined mesh that is computationally feasible for a given problem. We also assume that any mesh comprised of $N_{\text{fea}}+1$ uniformly spaced nodes in both the $x$ and $y$ directions, where $N_{\text{fea}}<N_{\text{max}}$ is computationally feasible. Given these assumptions, we can make the following recommendations:
\begin{itemize}
\item For $N_{\text{max}} > m$, the fine-scale solution model should be used with $N_f =N_{\text{max}}$;
\item For $N_{\text{max}} \leq m$, the coarse-scale model should be used with homogenization parameter $k = \lceil{m/N_{\text{max}}}\rceil$, periodic boundary conditions on the homogenization cells; and a mesh constructed by setting $N_{c} = m/k$.
\end{itemize}
We recommend that the fine-scale model be used if $N_{\text{max}} > m$ as the fine-scale solution does not require any homogenization. We recommend the use of periodic conditions in the case of $N_{\text{max}} \leq m$ for the following reasons:
\begin{itemize}
\item The periodic and confined boundary conditions have comparable \tpmsecondrevision{solution-based errors} that are significantly lower than the error for the uniform conditions across all tests, as seen in Figure \ref{fig:vf_50}.
\item The periodic and confined boundary conditions compute the correct effective diffusivities for layered media, whereas the uniform boundary conditions give an incorrect value for the effective diffusivity in the direction perpendicular to the layers \citep{szymkiewicz_2012}.
\item The implementation of the finite volume methods for the calculation of the effective diffusivities corresponding to homogenization cells with either periodic or uniform conditions requires the inversion of only one coefficient matrix, as the coefficient matrix $\mathbf{A}_{e}$ appearing in the linear system (\ref{eq:linear_system_ed}) is identical for both $\xi = x$ and $\xi = y$, whereas the linear system  that arises from confined conditions uses a different coefficient matrix for both $\xi = x$ and $\xi = y$ and thus requires two matrix inversions.
\item The periodic and uniform conditions are guaranteed to return symmetric positive definite effective diffusivities, which \tpmsecondrevision{ensures} that the principal axes of diffusion and the diffusivities in those directions are physical, whereas the confined conditions do \tpmsecondrevision{not} \citep{szymkiewicz_2012}.
\end{itemize}
We recommend $k = \lceil{m/N_{\text{max}}}\rceil$ as smaller values of $k$ are more accurate and it is the smallest value of $k$ giving a computationally-feasible mesh.
\section{Qualitative results}
\label{sec:qualitative_results}
\tpmsecondrevisionarticle{In this section we present some qualitative results to illustrate the effects of homogenization on steady-state diffusion problems}. As periodic boundary conditions on the homogenization cell were found to be the best-performing conditions and all four coarse-scale boundary conditions performed similarly, we consider only these periodic conditions on the homogenization cell and BC1 coarse-scale boundary conditions for the remainder of this paper.
\subsection{Homogenized solutions of steady-state diffusion problem with constant mesh}
\label{sec:steady_state}
We consider the solution of a steady-state diffusion problem on binary media with volume fraction $\varepsilon_1 = 50\%$, diffusivity ratio $\varepsilon_2 = 100$, $m = 60$, $N _c = m$  and \tpmrevision{$k = 1,3,5,10,15,30$}. Figure {\ref{fig:overlaid_axes_steady_state_periodic_BC_constant_mesh} shows these solutions as well as the original geometry with each principal axes of diffusion overlaid on each homogenization cell. For low levels of homogenization, $k = 3$ and $k = 5$, the solutions compare well to the fine-scale solution $k = 1$, although the contours representing $U = 0.1, 0.2$ and $0.3$ are significantly different in shape. For  $k = 10$, the majority of the solution field is signficantly different than that of the fine-scale solution, with the possible exception of the contours representing $U = 0.7, 0.8$ and $0.9$. For $k = 15$, the solution field bears only a vague resemblance to the fine-scale solution as the homogenized domain is comprised of only $16$ homogenization cells, indicating that a smaller value of $k$ is required to adequately capture the fine-scale behaviour. \tpmrevision{For $k = 30$ the solution is almost linear as there are only four different effective diffusivities across the entire domain.} 

\begin{figure}[htbp!]
\centering
\subfloat[{$k = 1$}]{\scalebox{1}[-1]{\includegraphics[height=0.28\textwidth]{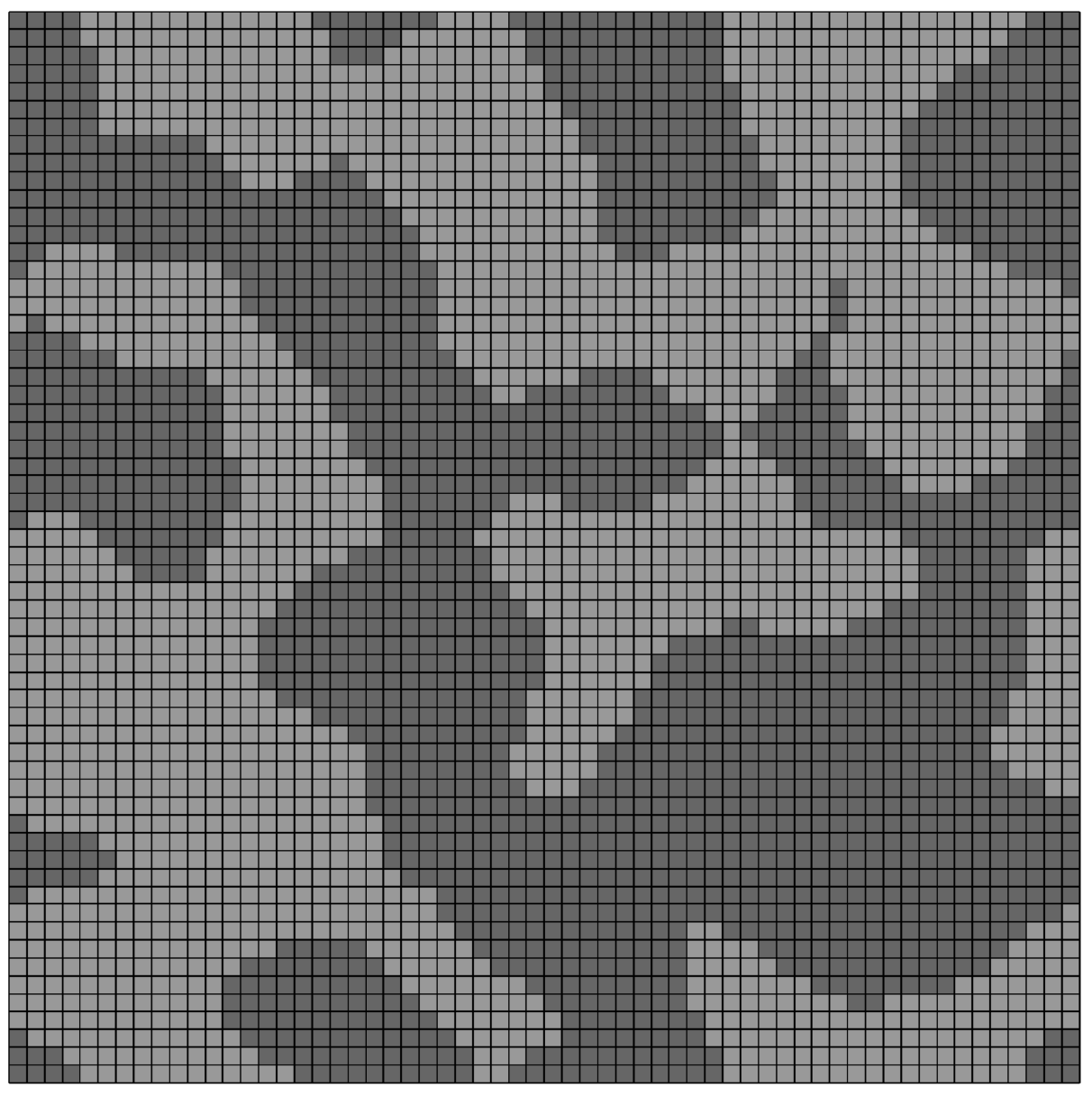}}}
\hspace{0.1cm}
\subfloat[{$k = 3$}]{\scalebox{1}[-1]{\includegraphics[height=0.28\textwidth]{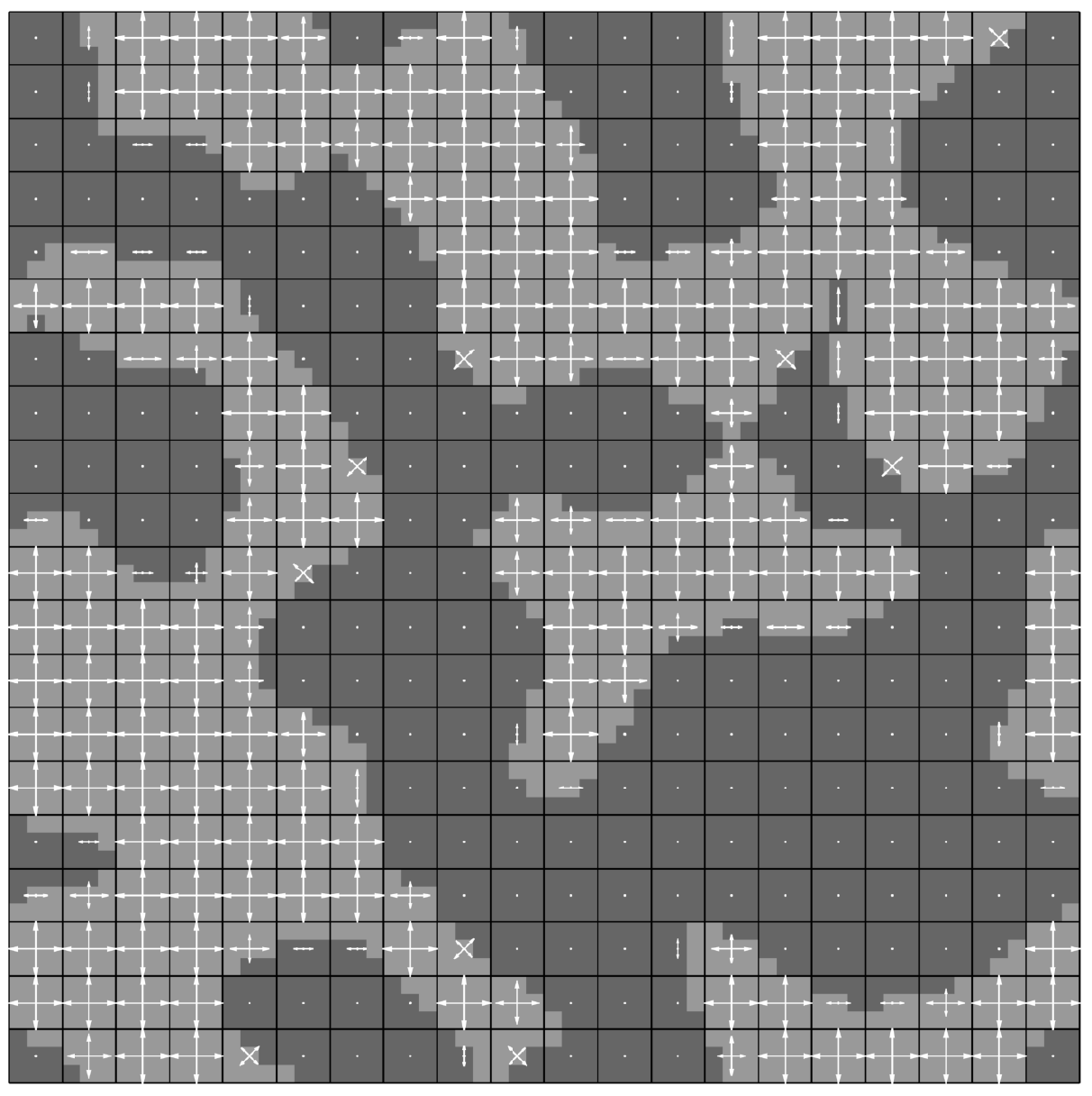}}}
\hspace{0.1cm}
\subfloat[{$k = 5$}]{\scalebox{1}[-1]{\includegraphics[height=0.28\textwidth]{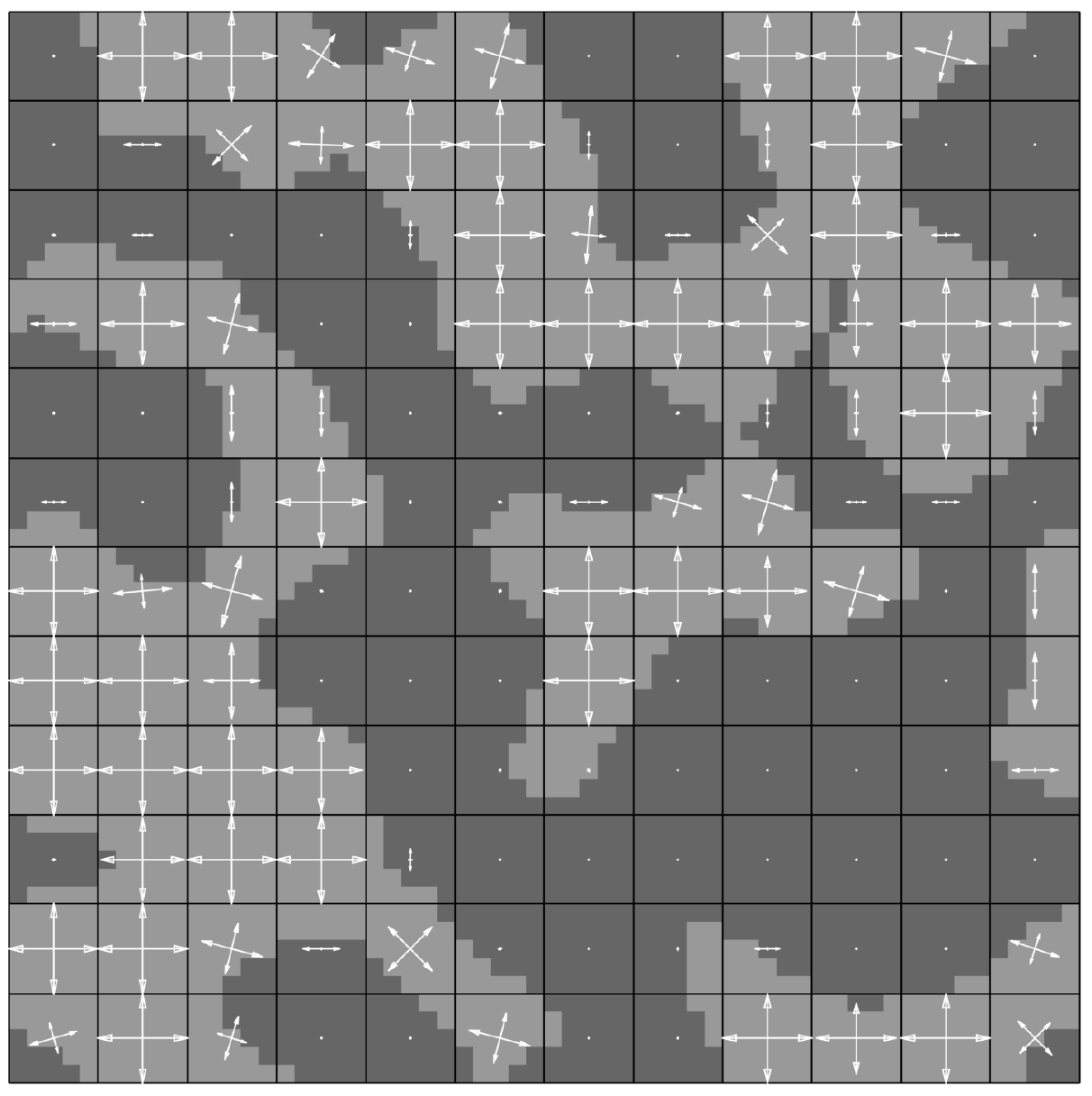}}}\\
\hspace{0.8cm}
\subfloat[{$k = 1$}]{\includegraphics[height=0.28\textwidth]{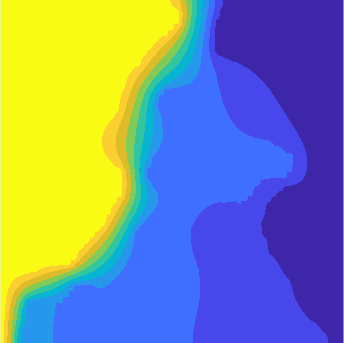}}
\hspace{0.1cm}
\subfloat[{$k = 3$}]{\includegraphics[height=0.28\textwidth]{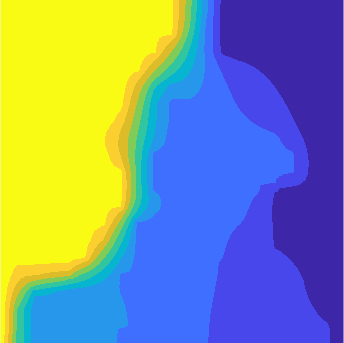}}
\hspace{0.1cm}
\subfloat[{$k = 5$}]{\includegraphics[height=0.28\textwidth]{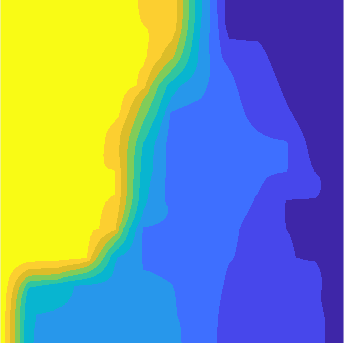}}
\hspace{0.1cm}
{\includegraphics[height=0.28\textwidth]{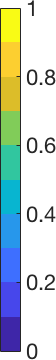}}\\
\subfloat[{$k = 10$}]{\scalebox{1}[-1]{\includegraphics[height=0.28\textwidth]{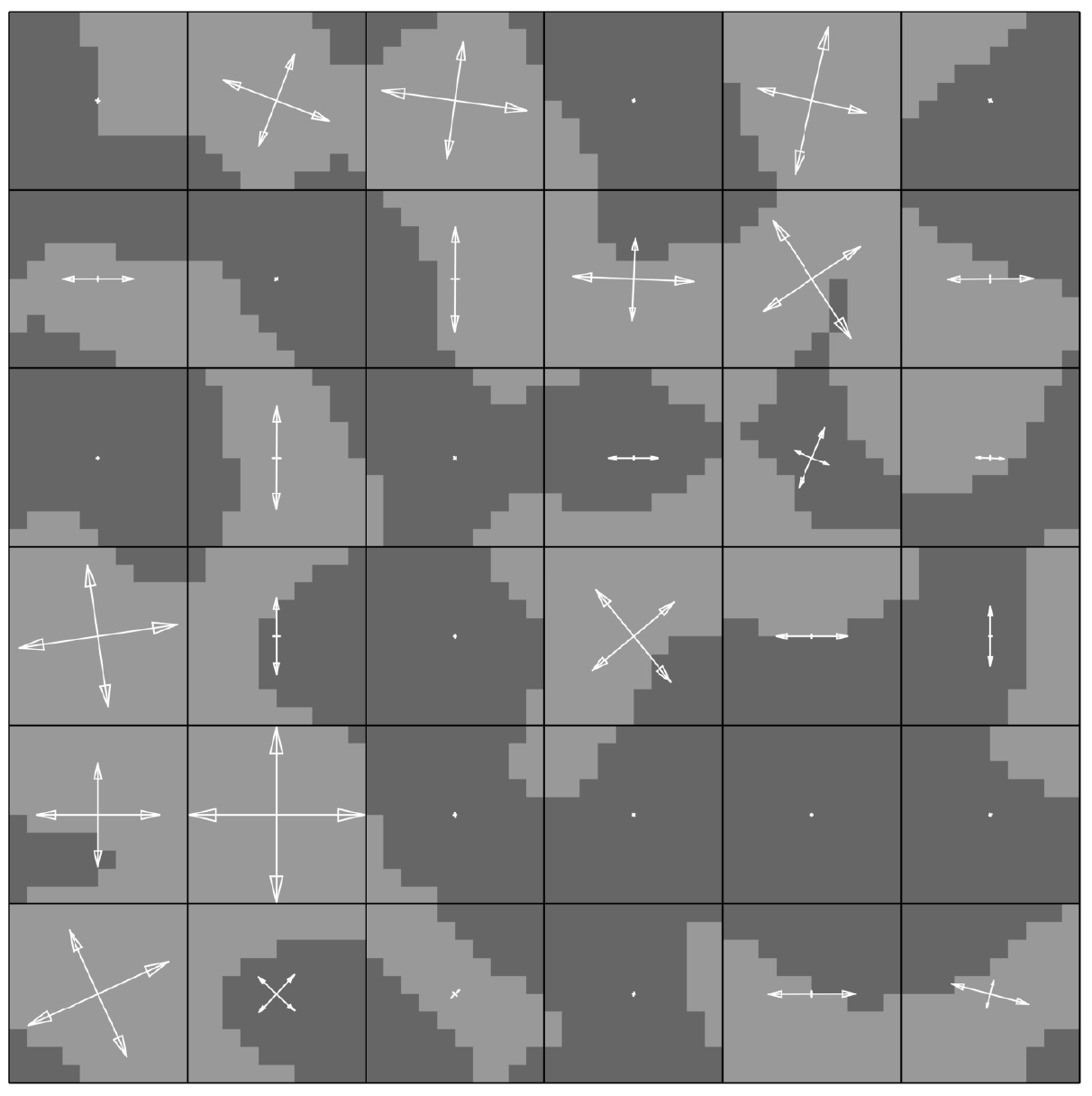}}}
\hspace{0.1cm}
\subfloat[{$k = 15$}]{\scalebox{1}[-1]{\includegraphics[height=0.28\textwidth]{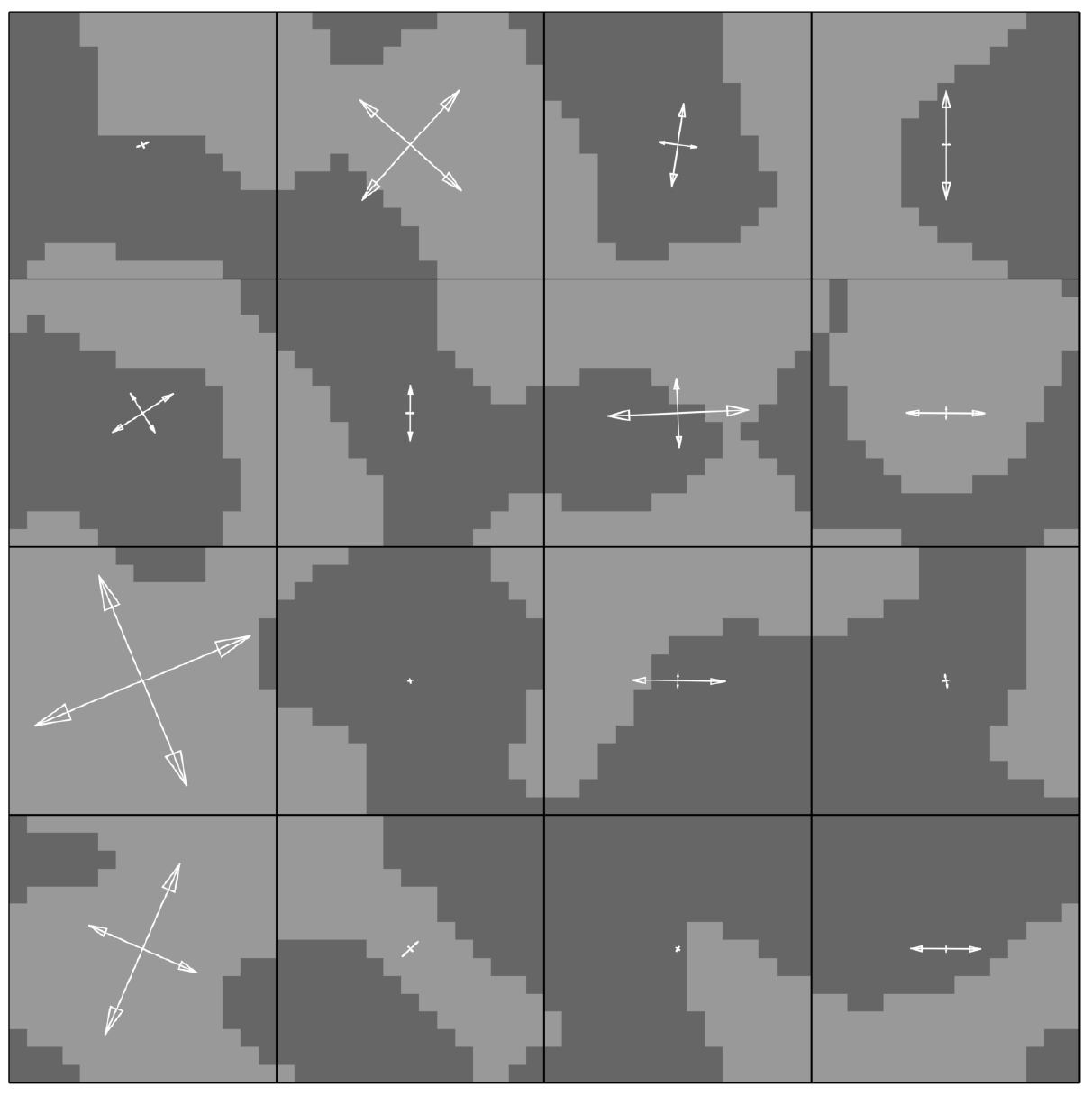}}}
\hspace{0.1cm}
\subfloat[{\tpmrevision{$k = 30$}}]{\scalebox{1}[-1]{\includegraphics[height=0.28\textwidth]{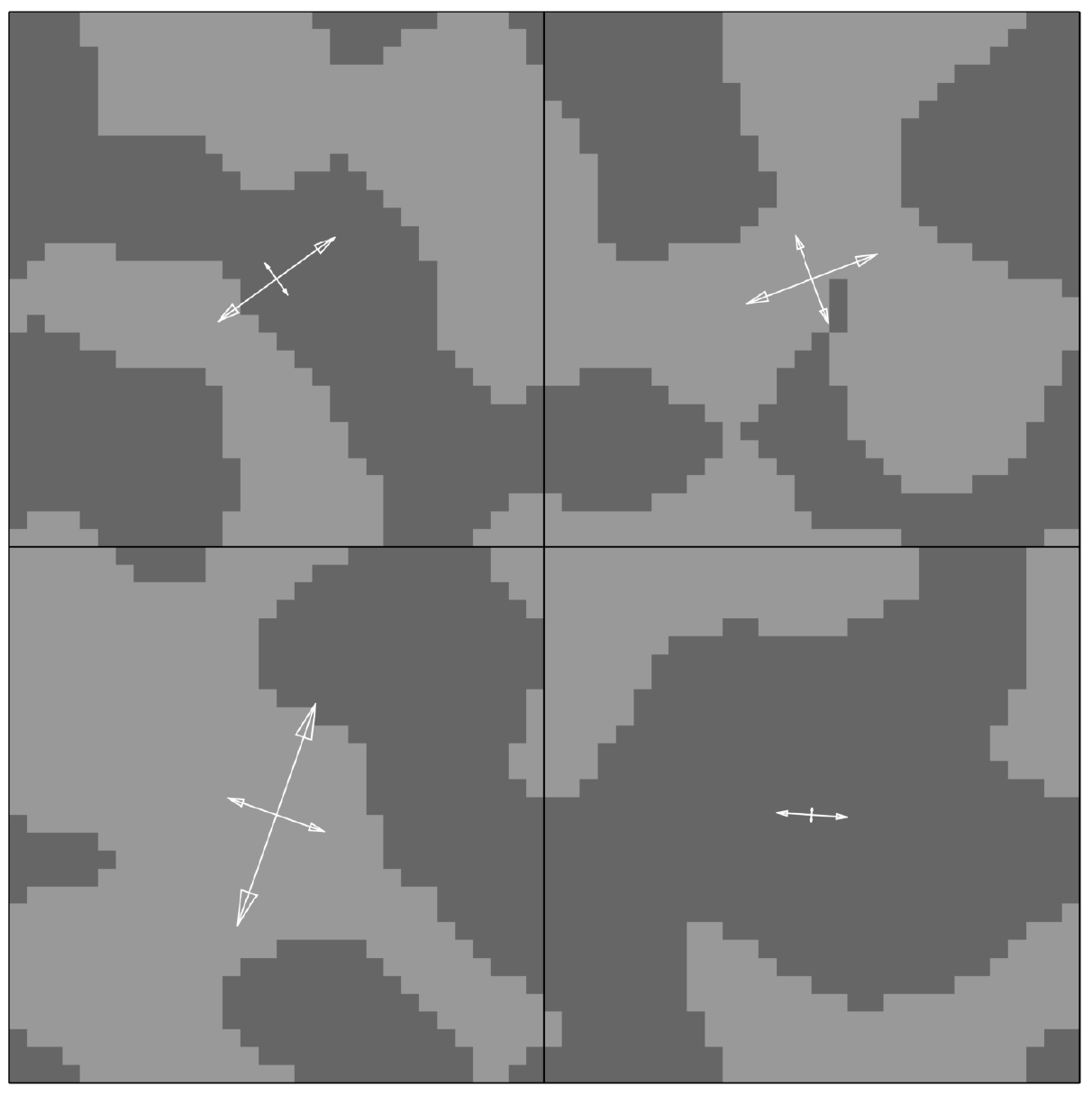}}}\\
\hspace{0.8cm}
\subfloat[{$k = 10$}]{\includegraphics[height=0.28\textwidth]{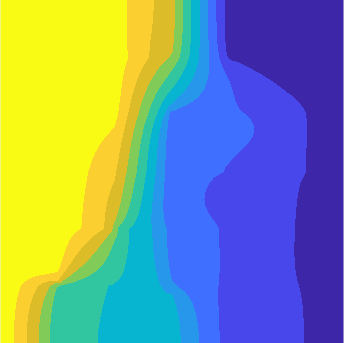}}
\hspace{0.1cm}
\subfloat[{$k = 15$}]{\includegraphics[height=0.28\textwidth]{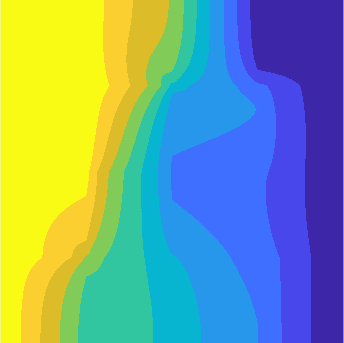}}
\hspace{0.1cm}
\subfloat[{\tpmrevision{$k = 30$}}]{\includegraphics[height=0.28\textwidth]{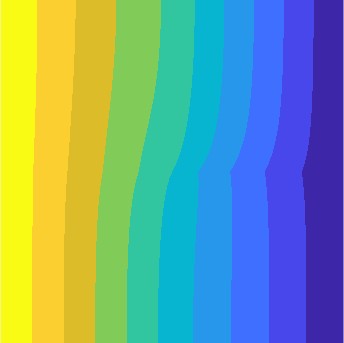}}
\hspace{0.1cm} 
{\includegraphics[height=0.28\textwidth]{Figures/Experiments_New/colorbar_v2}}\\
\caption{(a)--(c) and (g)--(k) Geometry with principal axes of diffusion overlaid for $k = 1,3,5,10,15,30$. The lengths of the axes have been scaled to reflect the diffusivity in the direction in which the axis points,  such that longer axes represent larger diffusivity values.  Dark grey indicates a diffusivity $D = 0.01$ and light grey indicates a diffusivity  $D = 1$. (d)--(f) and (j)--(l) Steady-state solution as calculated using a mesh with $N_c = 60$ for each of the homogenization parameters.}
\label{fig:overlaid_axes_steady_state_periodic_BC_constant_mesh} 
\end{figure}
  
\subsection{Homogenized solutions of steady-state diffusion problem with varying mesh}
\label{sec:steady_state_varying_mesh}
We repeat the experiments presented in section \ref{sec:steady_state} for $k = 2,3,6$ with a mesh constructed using $N_c = r$ as discussed in section \ref{sec:test2}. We consider only these homogenization parameters as we found that for larger values of $k$ that the corresponding mesh is too coarse to generate an accurate solution. We present these solutions in Figure \ref{fig:steady_state_periodic_BC_varying_mesh} and note that the appropriate comparison in this figure is between pairs of figures with the same values of $k$, as each solution is computed with the same homogenization parameter and the only difference is in the mesh used in the finite volume method to solve the coarse-scale model. For $k = 2$, the solutions are very similar, with the only noticeable difference in the solutions \tpmsecondrevision{occurring} in the contour representing $U = 0.4$ and $U = 0.3$. Similar results hold for $k = 3$, with the only noticeable difference in the contours of the solution \tpmsecondrevision{representing} $U = 0.4$ and $U = 0.3$. However for $k = 6$,  the solutions differ significantly, indicating that the mesh with $N_c = 10$ that is used in this simulation is too coarse to generate an accurate solution.

\begin{figure}
\centering
\subfloat[{$k = 2$}]{\includegraphics[height=0.28\textwidth]{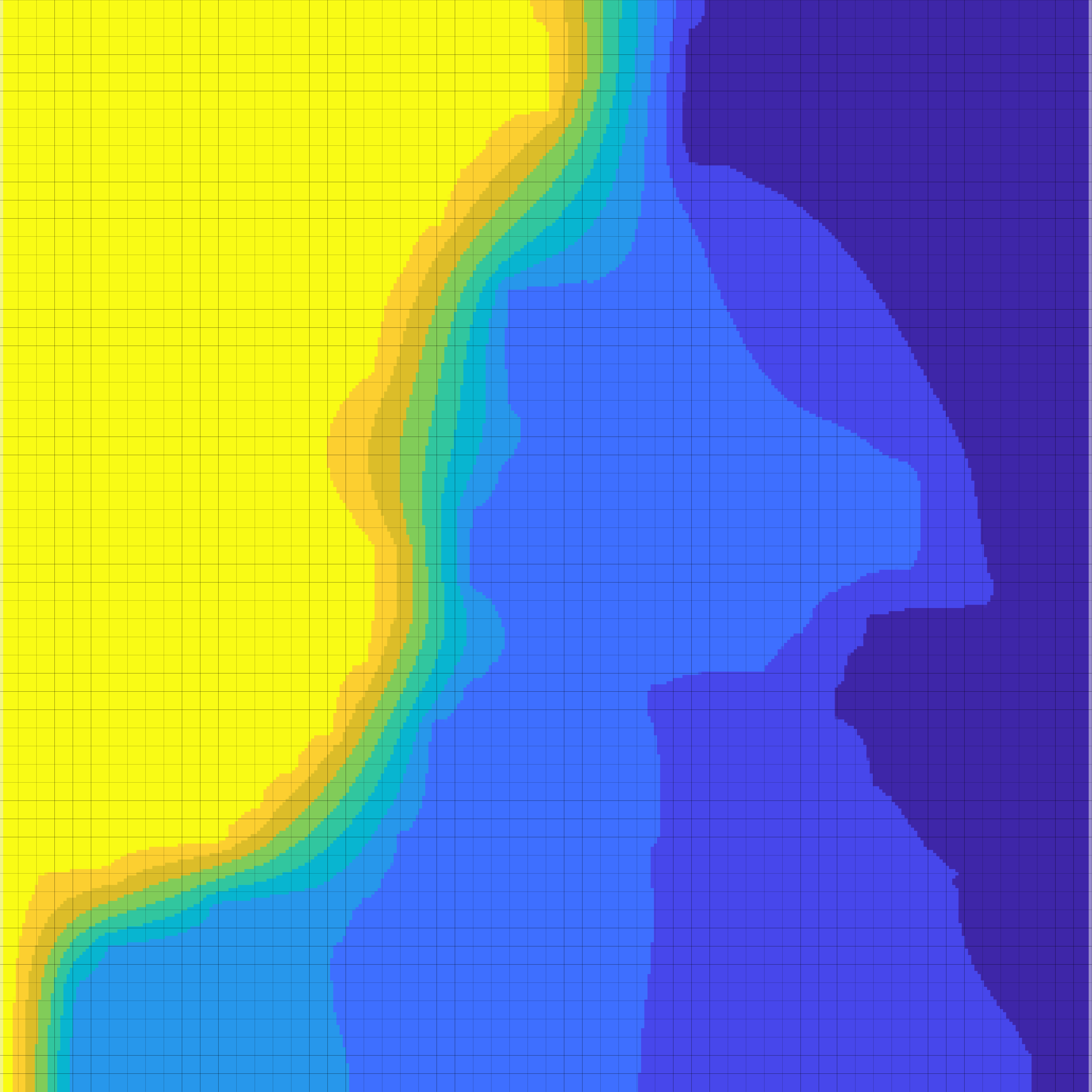}}
\hspace{0.1cm}
\subfloat[{$k = 3$}]{\includegraphics[height=0.28\textwidth]{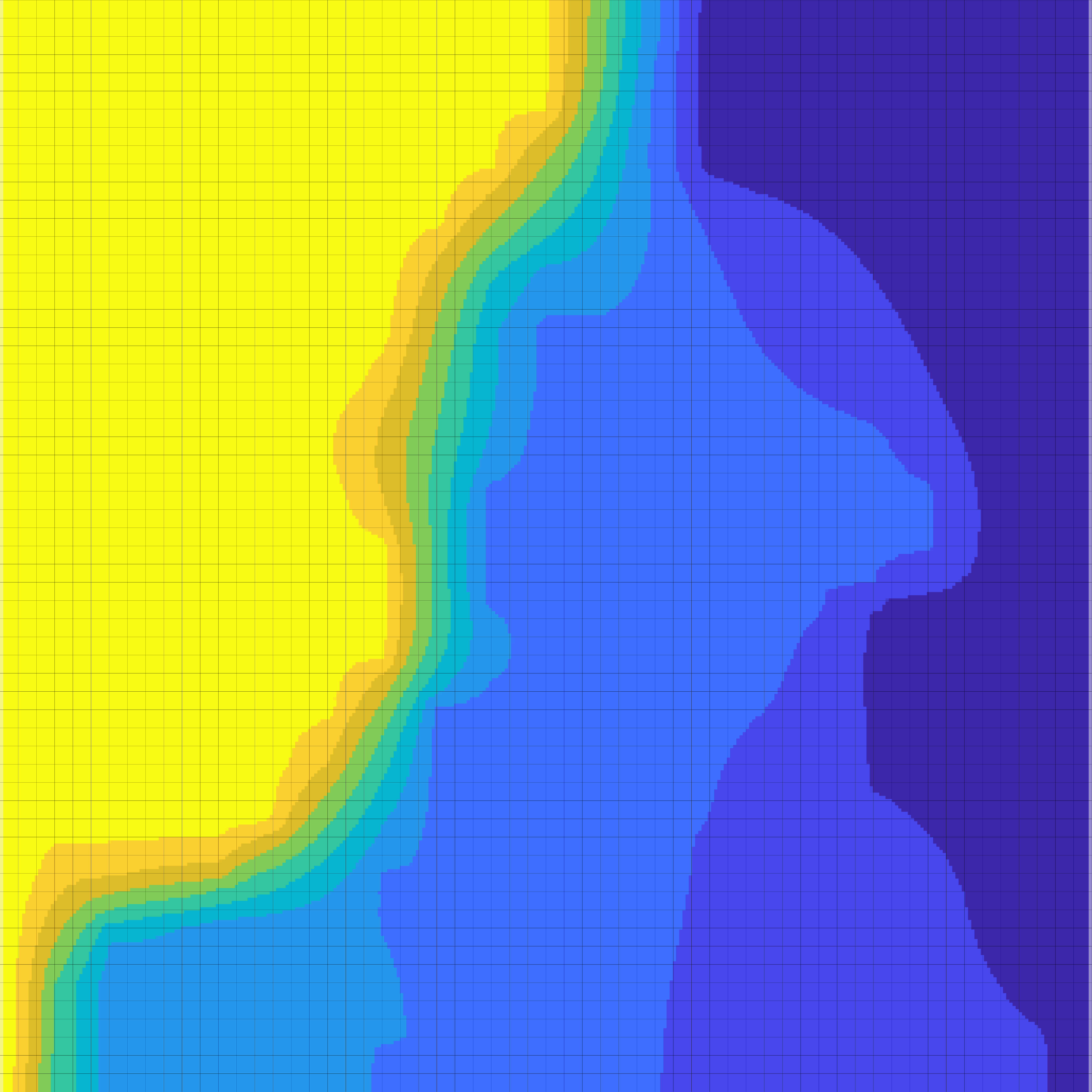}}
\hspace{0.1cm}
\subfloat[{$k = 6$}]{\includegraphics[height=0.28\textwidth]{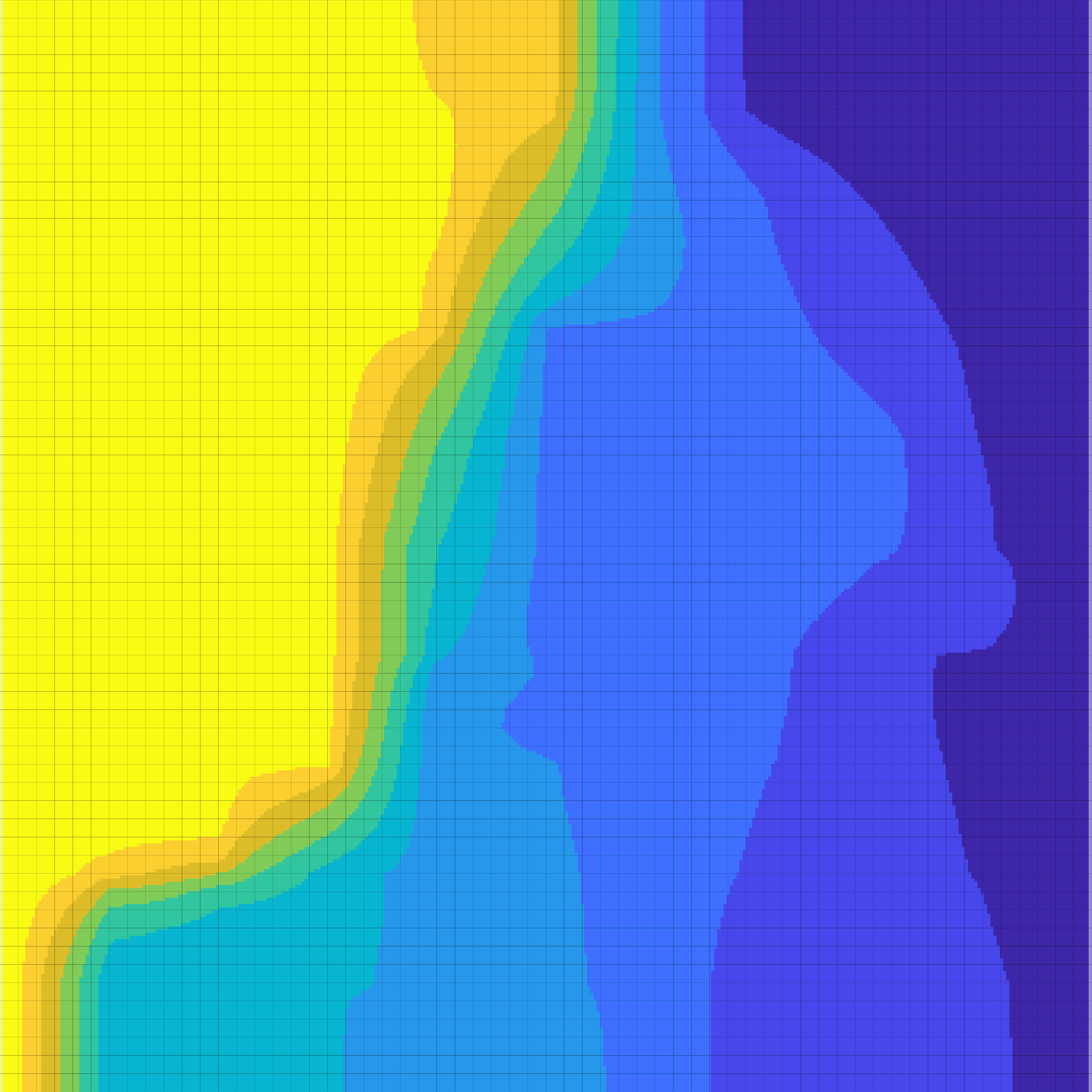}}
\hspace{0.1cm}
{\includegraphics[height=0.28\textwidth]{Figures/Experiments_New/colorbar_v2}}\\
\subfloat[{$k = 2$}]{\includegraphics[height=0.28\textwidth, width=0.28\textwidth]{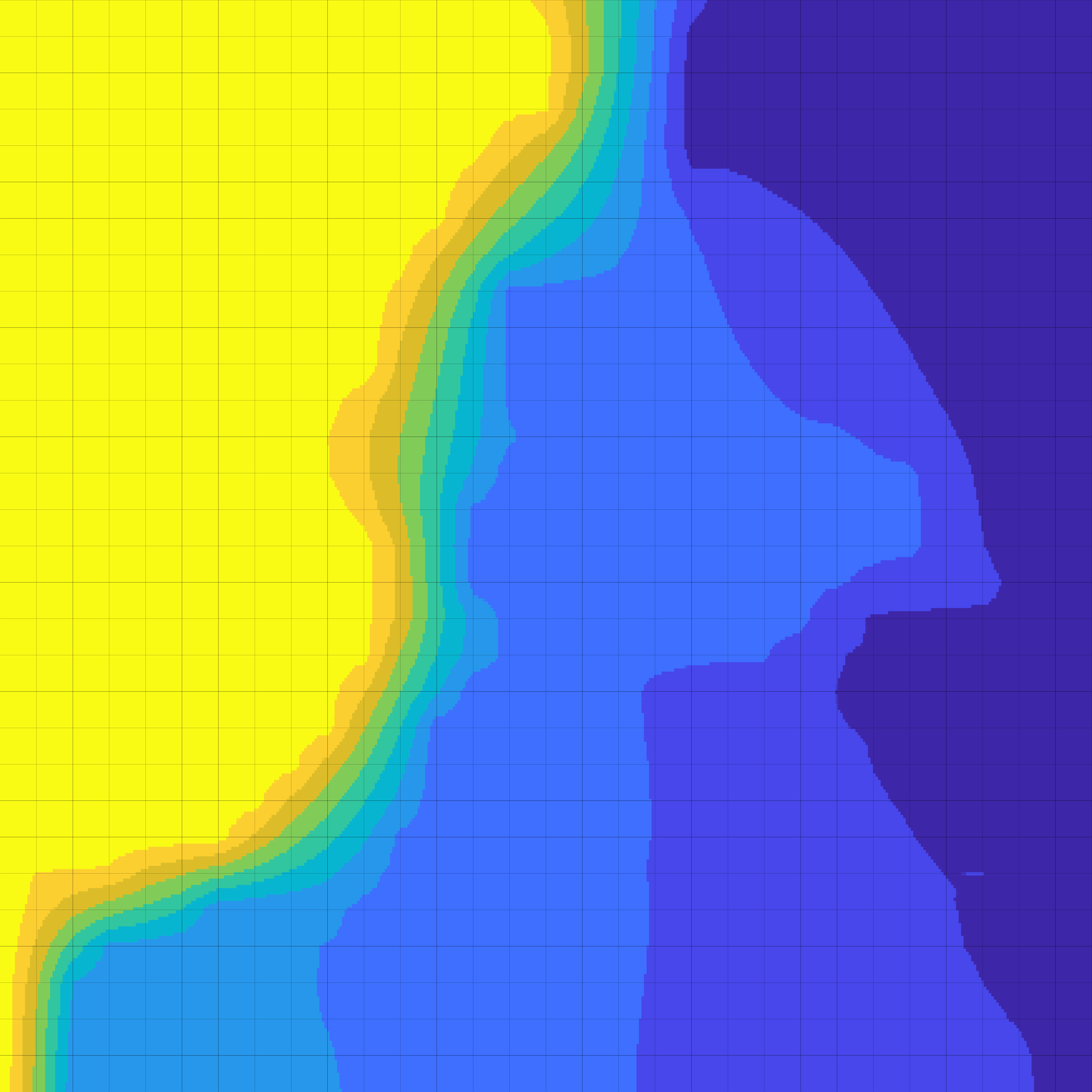}}
\hspace{0.1cm}
\subfloat[{$k = 3$}]{\includegraphics[height=0.28\textwidth, width=0.28\textwidth]{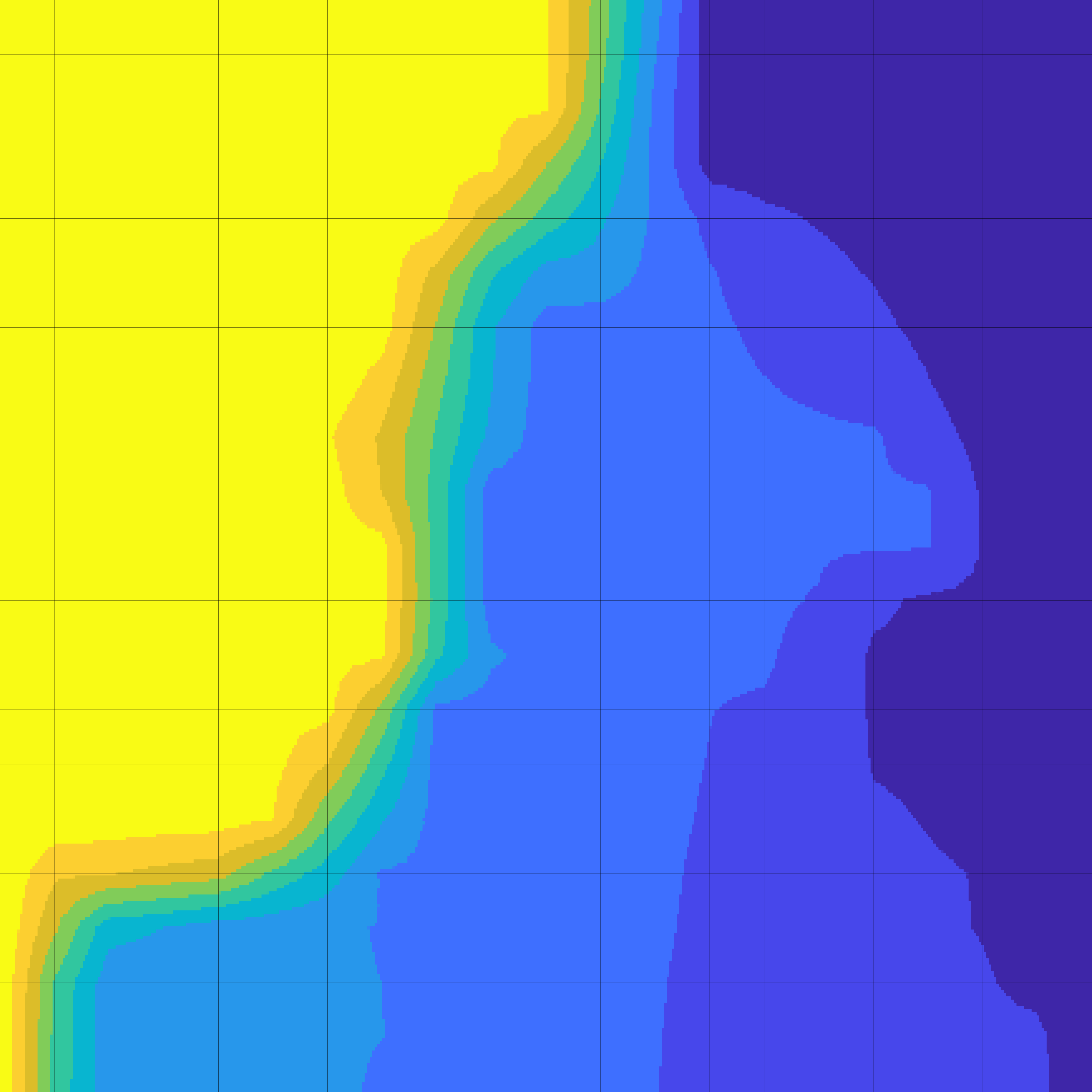}}
\hspace{0.1cm}
\subfloat[{$k = 6$}]{\includegraphics[height=0.28\textwidth, width=0.28\textwidth]{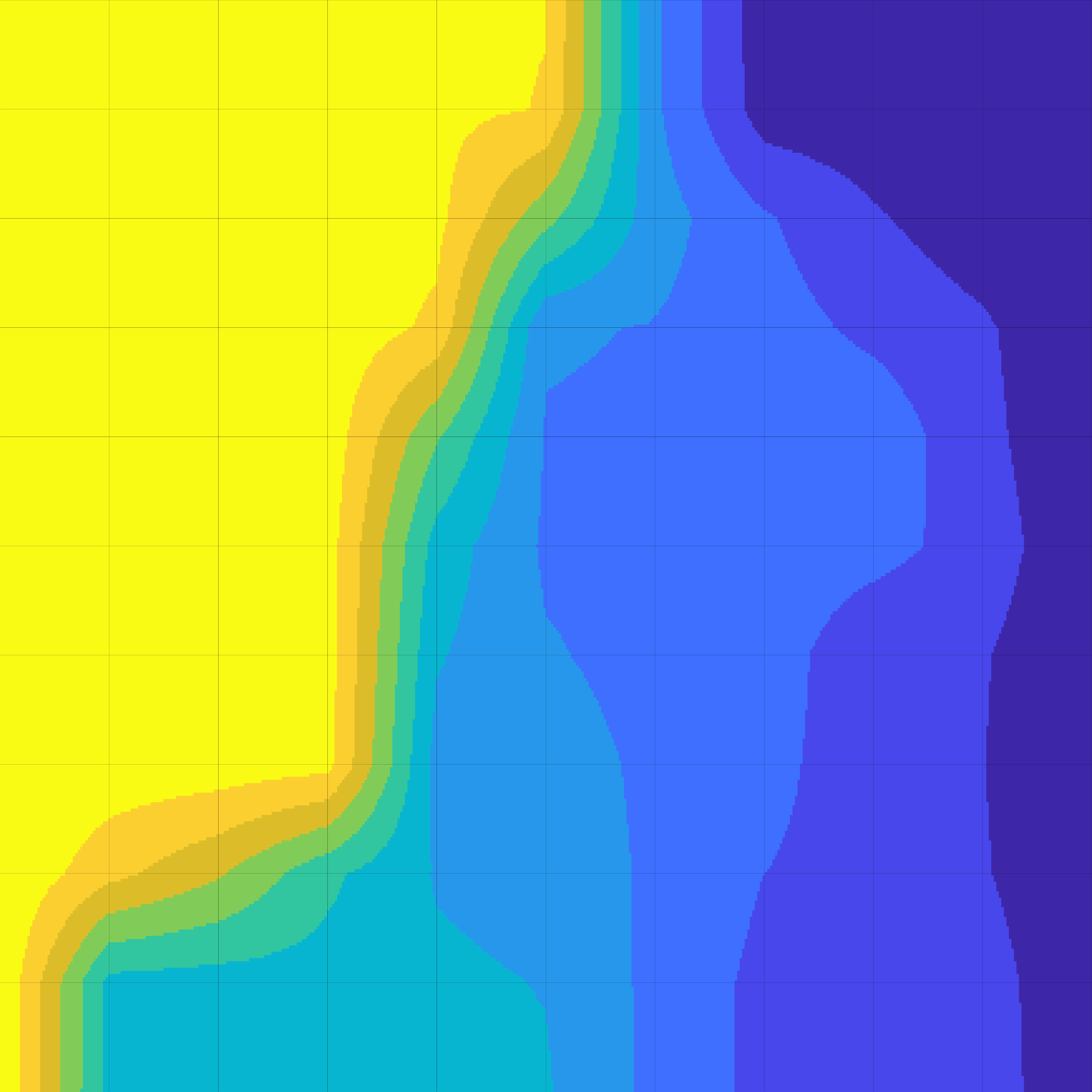}}
\hspace{0.1cm}
{\includegraphics[height=0.28\textwidth]{Figures/Experiments_New/colorbar_v2}}
\caption{(a)--(c) Steady-state solution as calculated using a mesh with $N_c = 60$ for $k = 2,3,6$. (d)--(f) Steady-state solution as calculated using a mesh with $N_c = r = m/k$ for $k = 2,3,6$. In all images the mesh used is overlaid upon the original image.}
\label{fig:steady_state_periodic_BC_varying_mesh}
\end{figure}

\tpmrevision{\subsection{Effects of minor geometry modifications on homogenization}
\label{sec:time_dependent_cg}
In this section, motivated by fracturing in shale gas reservoirs \citep{zhou_2017,kong_2018}, chalk engineering \citep{welch_2015,wattier_2018} and cell wall collapse in wood drying \citep{carr_2013b}, we compare coarse-scale models of diffusion on two different geometries, in which the second geometry is a slight modification of the first geometry. The purpose of this section is to investigate what effects these modifications have on the solution of steady-state diffusion problems and to what extent these changes can be seen in the coarse-scale solution. We consider a simulation in the media presented in Figure \ref{fig:changing_geometry_solutions}. In this geometry there are three sections that have been changed to generate the modified geometry (highlighted using red ellipses). In two of these changes a number of low diffusivity dark grey blocks are replaced with high diffusivity light grey blocks and in the other change one light grey block has been replaced with a dark grey block.  These first two changes allow for two continuous paths of light grey material joining the left and right boundaries of the domain while the other change has yielded a section of light grey material that is entirely surrounded by dark grey material. 

In the comparison of the fine-scale solutions represented in Figure \ref{fig:changing_geometry_solutions}(d) and Figure \ref{fig:changing_geometry_solutions}(j), the modifications to the geometry have caused the size of the regions between the contours corresponding to $u = 1$ and $u = 0.9$ and between $u = 0.1$ and $u = 0$ to decrease and the size of the other regions to increase. For the solutions corresponding to $k = 3$ (Figures \ref{fig:changing_geometry_solutions} (e) and (f)), they are both similar to the respective fine-scale solutions, with the largest difference occurring in the regions between the contours $U = 0.3$ and $U = 0.2$. For $k = 5$, there are significant differences in the region between the contours $U = 1$ and $U = 0.9$ for the coarse-scale solution Figure \ref{fig:changing_geometry_solutions}(l) compared to the fine-scale solution Figure \ref{fig:changing_geometry_solutions}(j), whereas differences corresponding to the comparison of the coarse-scale solution Figure \ref{fig:changing_geometry_solutions}(f) to the fine-scale solution Figure \ref{fig:changing_geometry_solutions}(d) are less pronounced. However, the reverse situation is true for the region between the contours $U = 0$ and $U = 0.1$. Overall, the process of homogenization has yielded similar qualitative differences in the coarse-scale solutions for both the original geometry Figure \ref{fig:changing_geometry_solutions}(a) and the modified geometry Figure \ref{fig:changing_geometry_solutions}(g). 

\begin{figure}[htbp!]
\vspace{-0.5cm}
\centering
\subfloat[{$k = 1$}]{\scalebox{1}[-1]{\includegraphics[height=0.28\textwidth, width=0.28\textwidth]{Figures/plot_periodic_geometry_1_BC_1_t1_constant_mesh_k_1}}}
\hspace{0.1cm}
\subfloat[{$k = 3$}]{\scalebox{1}[-1]{\includegraphics[height=0.28\textwidth, width=0.28\textwidth]{Figures/plot_periodic_geometry_1_BC_1_t1_constant_mesh_k_3}}}
\hspace{0.1cm}
\subfloat[{$k = 5$}]{\scalebox{1}[-1]{\includegraphics[height=0.28\textwidth, width=0.28\textwidth]{Figures/plot_periodic_geometry_1_BC_1_t1_constant_mesh_k_5}}}\\
\hspace{0.8cm}
\subfloat[{$k = 1$}]{\includegraphics[height=0.28\textwidth]{Figures/Experiments_New/plot_periodic_geometry_1_BC_1_k_1_constant_mesh_steady_state.png}}
\hspace{0.1cm}
\subfloat[{$k = 3$}]{\includegraphics[height=0.28\textwidth]{Figures/Experiments_New/plot_periodic_geometry_1_BC_1_k_3_constant_mesh_steady_state.png}}
\hspace{0.1cm}
\subfloat[{$k = 5$}]{\includegraphics[height=0.28\textwidth]{Figures/Experiments_New/plot_periodic_geometry_1_BC_1_k_5_constant_mesh_steady_state.png}}
\hspace{0.1cm}
{\includegraphics[height=0.28\textwidth]{Figures/Experiments_New/colorbar_v2}}\\
\subfloat[{$k = 1$}]{\scalebox{1}[-1]{\includegraphics[height=0.28\textwidth, width=0.28\textwidth]{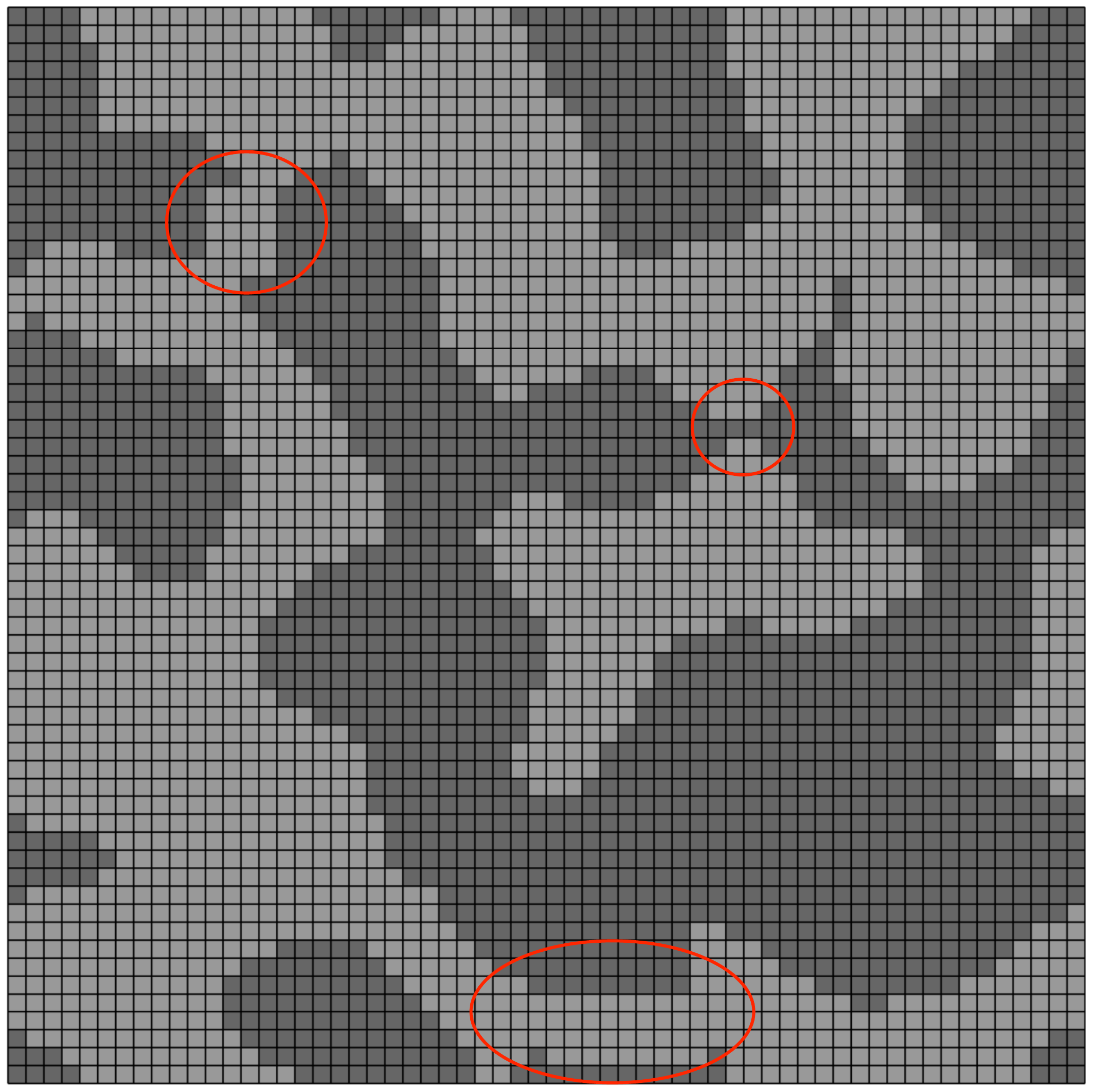}}}
\hspace{0.1cm}
\subfloat[{$k = 3$}]{\scalebox{1}[-1]{\includegraphics[height=0.28\textwidth, width=0.28\textwidth]{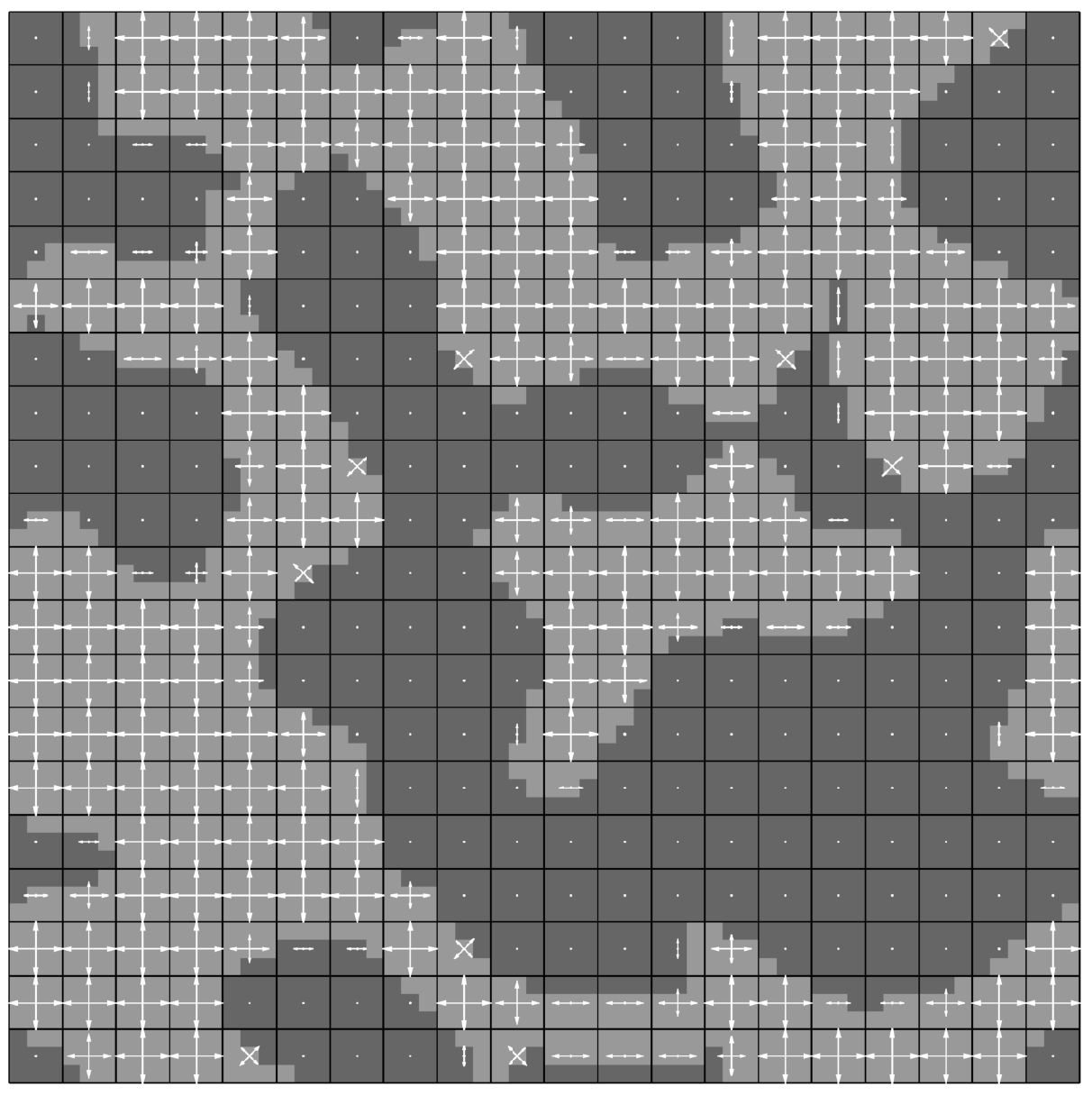}}}
\hspace{0.1cm}
\subfloat[{$k = 5$}]{\scalebox{1}[-1]{\includegraphics[height=0.28\textwidth, width=0.28\textwidth]{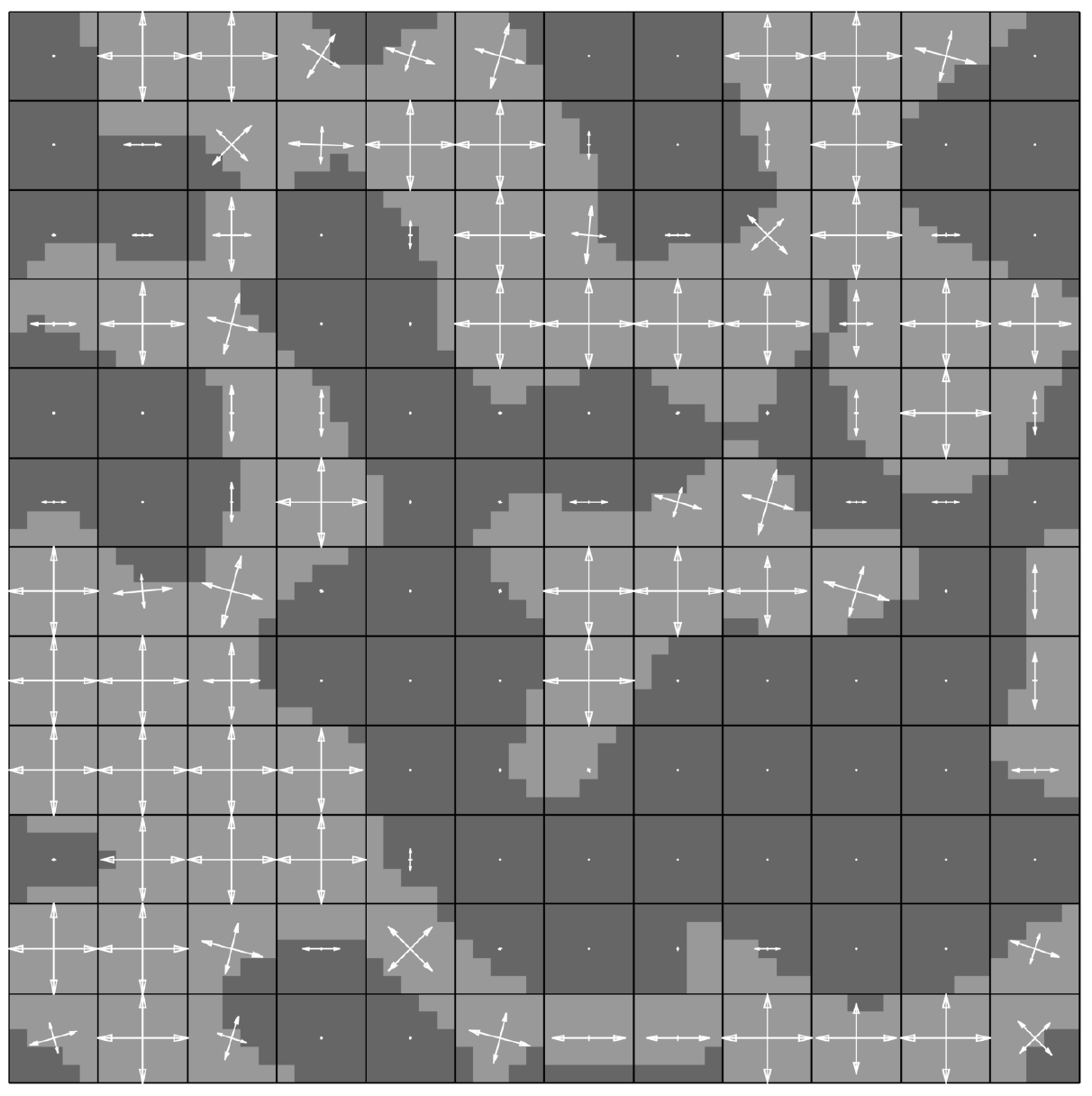}}}\\
\hspace{0.8cm}
\subfloat[{$k = 1$}]{\includegraphics[height=0.28\textwidth, width=0.28\textwidth]{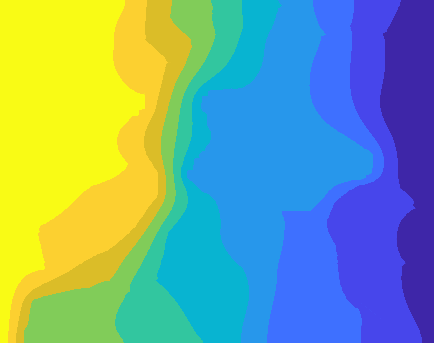}}
\hspace{0.1cm}
\subfloat[{$k = 3$}]{\includegraphics[height=0.28\textwidth, width=0.28\textwidth]{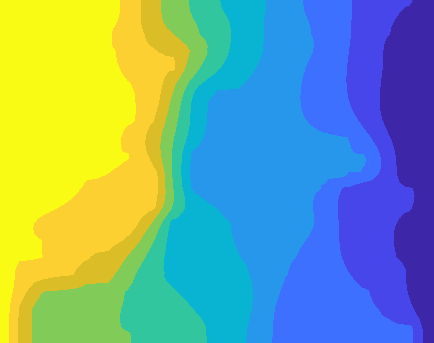}} 
\hspace{0.1cm}
\subfloat[{$k = 5$}]{\includegraphics[height=0.28\textwidth, width=0.28\textwidth]{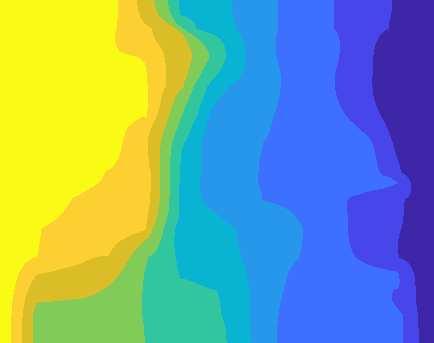}}
\hspace{0.1cm}
{\includegraphics[height=0.28\textwidth]{Figures/Experiments_New/colorbar_v2}}
\caption{(a)--(c) Geometry with principal axes of diffusion overlaid for $k = 1,3,5$. The lengths of the axes have been scaled to reflect the diffusivity in the direction in which the axis points,  such that longer axes represent larger diffusivity values.  Dark grey indicates a diffusivity $D = 0.01$ and light grey indicates a diffusivity  $D = 1$. (d)--(f) Solution at time $t = 0.5$ as calculated using a mesh with $N_c = 60$ for $k = 1,3,5$. (g)--(k) Changed geometry with principal axes of diffusion overlaid, with the changes indicated in (g) by red ellipses. (j)--(l) Solution at time $t = 1$ as calculated using a mesh with $N_c = 60$ for $k = 1,3,5$.}
\label{fig:changing_geometry_solutions}
\end{figure}

}\section{Conclusion}
\label{sec:conclusion}
\tpmsecondrevisionarticle{In this paper we have investigated homogenization techniques for steady-state diffusion problems on complex heterogeneous domains. We have considered periodic, uniform and confined boundary conditions on the homogenization cell and found that in general, periodic and confined conditions perform better than the uniform conditions with respect to the solution-based error and that all three types of boundary conditions performed similarly with respect to the flux-based error. Based on the outcomes of the study, we recommend periodic boundary conditions be chosen for the boundary value problem used in the homogenization of block heterogeneous media.} We also found that for steady-state problems the homogenization error is approximately $10$ to $50$ times larger than the spatial discretisation error.

We considered other factors that had no obvious impact on the accuracy of coarse-scale solutions, such as the boundary conditions on the coarse scale problem, different volume fractions of high diffusivity and low diffusivity material and the diffusivity ratio between the materials. \tpmsecondrevision{While different volume fractions and diffusivity ratios will affect the error that homogenization introduces, the choice of periodic conditions with the smallest computationally-feasible homogenization parameter is recommended for simulating diffusion in complex heterogeneous media}. We note that our analysis is limited to Dirichlet and Neumann conditions on the boundaries as we wished to minimise the impact of the coarse-scale boundary conditions on the solution. We considered only binary media in this work however the methods could be applied to media which consist of an arbitrary number of different diffusivities.

\section*{Acknowledgments}
The second and third authors acknowledge funding from the Australian Research Council (DE150101137, DP150103675). All authors acknowledge the helpful comments of the anonymous reviewers and editors that helped improve the quality of the manuscript.

\bibliographystyle{spbasic}      
\bibliography{mypapers}

\end{document}